\documentclass[apj]{emulateapj}
\usepackage{apjfonts}
\shorttitle{The BGPS Catalog}
\shortauthors{Rosolowsky et al.}

\newcommand{\ubc}{1}
\newcommand{\utexas}{2}
\newcommand{\casa}{3}
\newcommand{\ucf}{4}
\newcommand{\penn}{5}
\newcommand{\wisc}{6}
\newcommand{\jpl}{7}
\newcommand{\virginia}{8}
\newcommand{\hawaiihilo}{9}
\newcommand{\hawaiimain}{10}

\begin{document}

\title{The Bolocam Galactic Plane Survey --
  II. Catalog of the Image Data}

\author{Erik Rosolowsky\altaffilmark{\ubc},
Miranda K. Dunham\altaffilmark{\utexas},
        Adam Ginsburg\altaffilmark{\casa},
    Eric Todd Bradley\altaffilmark{\ucf},
James Aguirre\altaffilmark{\penn},
        John Bally\altaffilmark{\casa}, 
    Cara Battersby\altaffilmark{\casa},
    Claudia Cyganowski\altaffilmark{\wisc},
    Darren Dowell\altaffilmark{\jpl}
    Meredith Drosback\altaffilmark{\virginia},
    Neal J. Evans II\altaffilmark{\utexas},
        Jason Glenn\altaffilmark{\casa},
        Paul Harvey\altaffilmark{\utexas,\casa},
        Guy S. Stringfellow\altaffilmark{\casa},
        Josh Walawender\altaffilmark{\hawaiihilo}, 
        Jonathan P. Williams\altaffilmark{\hawaiimain}
}

\affil{{$^\ubc$}{University of British Columbia Okanagan, 3333 University Way,
  Kelowna BC, V1V 1V7, Canada}}
\email{\tt erik.rosolowsky@ubc.ca}


\affil{{$^\utexas$}{\it{Department of Astronomy, University of Texas,
      1 University Station C1400, Austin, TX 78712}}}

\affil{{$^\casa$}{\it{CASA, University of Colorado, 389-UCB, Boulder, CO 80309}}}

\affil{{$^\penn$}{\it{Department of Physics and Astronomy, University of
      Pennsylvania, Philadelphia, PA }}}

\affil{{$^\ucf$}{\it{Department of Physics, University of Central
Florida}}}

\affil{{$^\wisc$}{\it{Department of Astronomy, University of Wisconsin,
       Madison, WI 53706}}}

\affil{{$^\jpl$}{\it{Jet Propulsion Laboratory, California Institute
   of Technology, 4800 Oak Grove Dr., Pasadena, CA 91104}}}

\affil{{$^\virginia$}{\it{Department of Astronomy, University of
Virginia, P.O. Box 400325, Charlottesville, VA 22904}}}



\affil{{$^{\hawaiihilo}$}{\it{Institute for Astronomy, 640 N. Aohoku Pl., Hilo, HI 96720}}}

\affil{{$^{\hawaiimain}$}{\it{Institute for Astronomy,2680 Woodlawn Drive, Honolulu, HI 96822}}}

%

\begin{abstract}
We present a catalog of 8358 sources extracted from images produced by
the Bolocam Galactic Plane Survey (BGPS).  The BGPS is a survey of the
millimeter dust continuum emission from the northern Galactic plane.
The catalog sources are extracted using a custom algorithm, Bolocat,
which was designed specifically to identify and characterize objects
in the large-area maps generated from the Bolocam instrument.  The
catalog products are designed to facilitate follow-up observations of
these relatively unstudied objects. The catalog is 98\% complete from
0.4 Jy to 60 Jy over all object sizes for which the survey is
sensitive ($<3.5'$). We find that the sources extracted can best be
described as molecular clumps -- large dense regions in molecular
clouds linked to cluster formation.  We find the flux density
distribution of sources follows a power law with $dN/dS\propto
S^{-2.4\pm 0.1}$ and that the mean Galactic latitude for sources is
significantly below the midplane: $\langle b \rangle=(-0.095\pm
0.001)^{\circ}$.
\end{abstract}

\keywords{Galaxies:Milky Way --- ISM:star formation}

\section{Introduction}
\label{intro}

The advent of large-number bolometer arrays at good observing sites
has enabled the mapping of large sections of the sky in the millimeter
continuum.  These observations are ideally suited for unbiased
searches for dense gas traced in dust thermal continuum emission.
Even at the cold temperatures and high densities found in molecular
clouds, dust emission remains an optically thin tracer of the gas
column density.  Despite some uncertainties in the properties of dust
emission in these regimes, these observations represent the most
efficient way to detect high column density features in the molecular
interstellar medium (ISM).  The original surveys of (sub)millimeter
continuum emission focused on nearby molecular clouds
\citep[e.g.,][]{motte-andre, johnstone-scuba,enoch07}.  With
improvements in instrumentation and methodology, surveys of large
sections of the Galactic plane have become possible.  This paper
presents results from one such survey: the Bolocam Galactic Plane
Survey (BGPS).  The Bolocam instrument at the Caltech Submillimeter
Observatory has been used to conduct a survey of the Galactic plane
from $-10^{\circ}<\ell < 90.5^{\circ}$ and $-0.5^{\circ} <b
<0.5^{\circ}$ in the continuum at $\lambda=1.1$~mm.  The survey has
been augmented with larger latitude coverage from $77^{\circ}<\ell <
90.5^{\circ}$ and studies of individual star forming regions near
$\ell = 111^{\circ}$ and in the W3/4/5, Gem OB1 and IC 1396 regions.
The observations, reduction and calibration of survey data are
presented in a companion paper by \citet{bgps-paper1}, hereafter Paper
I.  This paper presents the catalog of millimeter continuum sources in
the surveyed region.

The interpretation of millimeter continuum maps from nearby molecular
clouds is guided by preceding studies in an array of molecular gas
tracers that developed a detailed understanding of the density
structure of clouds \citep[see the review by ][]{evans-araa}.  The
millimeter continuum structures detected in the observations of nearby
molecular clouds were molecular cores, i.e., dense structures that are
linked to the formation of individual stellar systems.  The detection
of cores was driven by sensitivity to bright objects with angular
scales of tens of arcseconds and filtering out larger angular scale
structures.

Although the sensitivity and filtering properties of the instruments
remain the same when directed at the Galactic plane, the typical
distance to the objects becomes significantly larger ($\sim$5 kpc
vs. $\sim$250 pc).  In addition, the range of possible distances spans
foreground objects in nearby star forming regions to bright features
viewed from across the Galactic disk.  Hence, the physical nature of
the objects seen in millimeter surveys of the Galactic plane is less
clear than it is for nearby molecular clouds.  Guided again by line
studies of nearby molecular clouds, surveys such as the BGPS are
likely detecting a range of structures within the molecular ISM.  The
actual objects that are detected are a product of matching the average
column density of dust to the angular scales over which the survey is
sensitive.  Given the distance to the objects, many different types of
objects may have an average column density detectable on the $\sim
30''$ scales to which the survey is most sensitive.

In addition to the BGPS, other surveys are beginning to explore the
Galactic plane in the millimeter dust continuum.  The ATLASGAL survey
uses the Laboca instrument on the APEX telescope to survey the
southern Galactic plane \citep{atlasgal} at $\lambda=0.87$~mm.  After
commissioning, the SCUBA2 instrument will be used to survey the plane
at $\lambda=0.85$~mm in the JCMT Galactic Plane Survey (JPS)
\citep{jps}.  These data will complement the open time Hi-GAL project
of the Herschel Space Telescope which will observe the inner Galactic
plane in several far infrared bands \citep{higal}.  The BGPS and these
other surveys are conducted in a {\it blind} fashion, meaning that the
survey spans all of a large region with no preferential targeting of
subregions within that section.  Since the survey regions span the
low-latitude Galactic plane, they likely cover a large fraction of the
star forming regions in the Galaxy.  Hence, the maps will serve as the
finding charts for studying dense gas in our Galaxy using the next
generation of instrumentation, in particular the Atacama Large
Millimeter Array.

This paper presents a catalog of the BGPS data suitable for
statistical analysis and observational follow-up of the millimeter
objects in the Galactic plane.  The catalog is produced by a
tailor-made algorithm, Bolocat, which was designed and tested in
concert with the data reduction and calibration of the BGPS images.
Consequently, the algorithm specifically reflects and accounts for the
peculiarities in the BGPS images in a fashion that a generic source
extraction algorithm could not.  We present the catalog through a
brief description of the BGPS data (\S\ref{bgpssum}) and Bolocat
(\S\ref{howto}).  We present our validation of the source extraction
algorithm in \S\ref{testing} and the catalog itself in
\S\ref{thecatalog}.  Using observations of the NH$_3$ inversion
transitions towards several of the objects, we are able to present
constraints on the physical properties of catalog sources which are
discussed in \S\ref{physical}.  Finally, we present several summary
products from the catalog in \S\ref{summaryprod}.

\section{The Bolocam Galactic Plane Survey}
\label{bgpssum}
The BGPS is described in full detail in Paper I.  The survey uses the
144-element Bolocam array on the Caltech Submillimeter Observatory,
which observes in a band centered at 268 GHz (1.1 mm) and a width of
46 GHz \citep{bolocam-spie}.  Of particular note, the bandpass is
designed to reject emission from the CO($2\to 1$) transition, which is
the dominant line contributor at these wavelengths.  The survey covers
150 square degrees of the Galactic plane, primarily in the first
Galactic quadrant.  In the first quadrant, the survey ranges from
-0.5$^{\circ}$ to 0.5$^{\circ}$ in latitude.  Survey data are observed
in perpendicular scans made in the longitude and latitude direction,
with the exception of one block of scans in the W5 region.  All of the
survey fields are observed in either $1^{\circ}\times 1^{\circ}$ or
$3^{\circ}\times 1^{\circ}$ sub-fields which are then jointly reduced
producing several individual maps of the plane spanning several
degrees.

Imaging of the time stream data requires correcting the signal for the
dominant atmospheric contribution in the bolometer array.  Since the
fundamental assumption in the correction is that the atmosphere is
constant across the array, the data are not sensitive to spatial
structures with sizes larger than the array ($7.5'$).  More detailed
analysis shows that structures larger than $3.5'$ are not fully
recovered.  An astrophysical image is generated by analyzing the
variations of the signal in the time domain and identifying structures
due to astrophysical signal in the image domain.  By iteratively
refining the estimates of instrumental and atmospheric contributions
to the time stream, these effects can be decoupled from astrophysical
signal producing a map containing only signal and irreducible noise.
Of note for the cataloging process, the mapping procedure yields (1)
the BGPS map (2) a model of the signal at each position or
equivalently the residual of subtracting the model from the survey map
and (3) a record of the number of times a bolometer sampled a given
position on the sky.  The pixel scale of the BGPS is set to $7.2''$ or
4.58 pixels across the $33''$ FWHM beam.

\section{Catalog Methodology}
\label{howto}

\begin{figure*}
\plotone{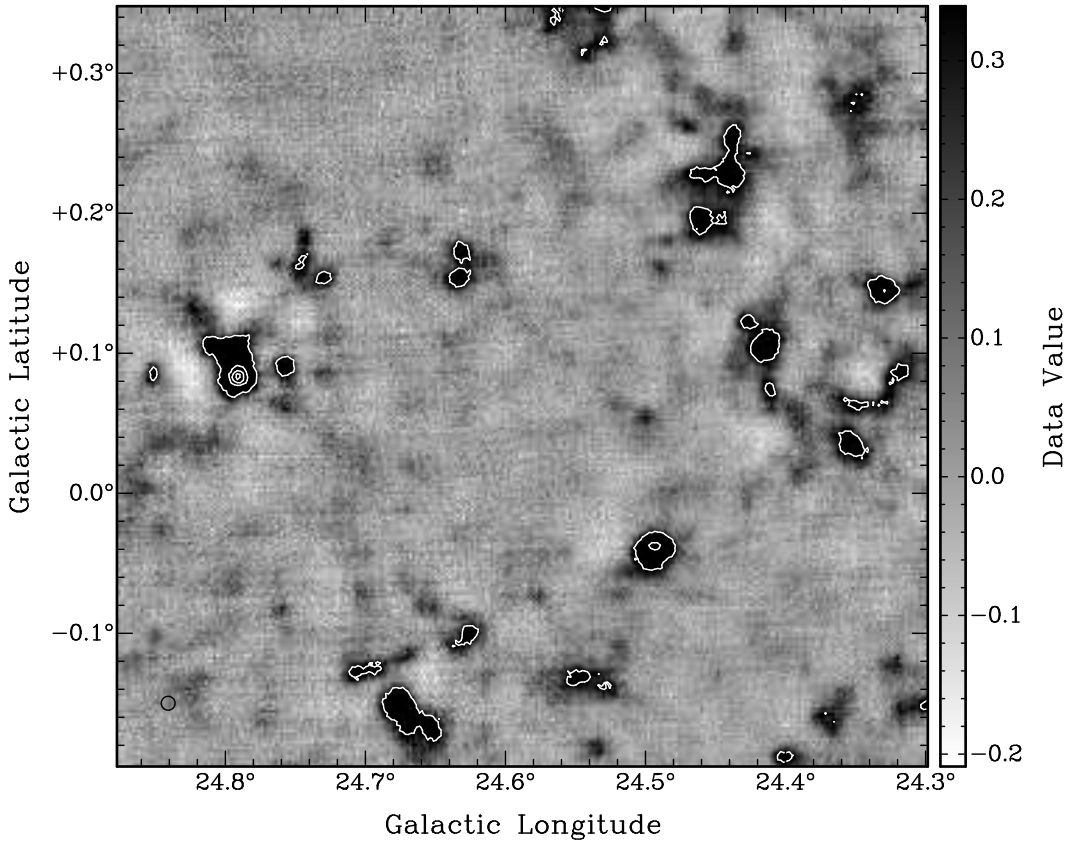}
\caption{\label{bgpssamp} A sample subsection from the BGPS map near
  $\ell=24^{\circ}$ where the units of the image are Jy beam$^{-1}$.
  White contours indicate the shapes of features above the top end of
  the grayscale, beginning at 0.33 Jy beam$^{-1}$ and increasing by
  0.33 Jy beam$^{-1}$ linearly.  The field highlights several of the
  structures that informed the design of the cataloging algorithm,
  including bright compact sources and filamentary structures as well
  as spatially varying noise and negative bowls around strong
  sources.}
\end{figure*}

The BGPS catalog algorithm was developed jointly with the mapping
process in response to the goals of the BGPS.  The BGPS is a
large-scale survey of the Galactic plane in a comparatively unexplored
waveband.  The resulting images contain a wealth of structure as
presented in Paper I.  A sample image appears in Figure \ref{bgpssamp}
displaying many of the features that motivate the structure of the
catalog algorithm.  The map contains several bright, compact sources
which frequently appear in complexes.  The bright sources commonly
have low intensity skirts or filaments of fainter emission surrounding
or connecting them.  In addition, both bright and faint sources have
resolved, irregular shapes.  The brightest sources commonly have
negative bowls around them as a consequence of the PCA cleaning during
the iterative mapping process (Paper I).  These bowls can affect flux
density measurements for these objects and depress the significance of
surrounding emission.  The maps also show noise with a spatially
varying rms on scales from pixel (sub beam) to degree scales.  Some
artifacts that have not been completely removed by the imaging process
still remain, manifesting as correlated noise in the scan directions.

Given these features in the images, we opted to design our own
cataloging algorithm rather than adapt a preexisting routine to the
BGPS data such as Clumpfind \citep{clumpfind} or Source EXtractor
\citep{sextractor}.   Since the physical nature of the objects is not
well-constrained, we did not impose prior information on the objects
that were to be extracted, such as the functional form of the
intensity profile.  Instead, the algorithm separates emission into
objects defined by a single compact object and nearby emission.  The
catalog is driven by the need to recover and characterize the
properties of the compact sources.  The data retained in the catalog
needed to be well-suited for follow-up observations, directing
observers to the most interesting and bright features in the images.
However, filamentary structure commonly appears without an associated
compact source and we developed a routine that could also characterize
isolated filaments of emission as objects.  The routine also needed to
account for the non-uniform nature of the noise properties over the
maps.  The algorithm described below is the result of applying these
design criteria.

We investigated the behavior of both Clumpfind and Source EXtractor on
a subset of the data to verify that a new approach was necessary.  Our
algorithm shares a common design philosophy with these other
approaches.  For the recommended user parameters, the Clumpfind
algorithm tended to break regions into sub-beam sized objects based on
small scale features in the objects from both real source structure
and the spatially varying noise.  After exploring the input parameters
to Source EXtractor, we were unable to satisfactorily identify
filamentary structure in the images.  Furthermore, those bright
objects that were detected were not divided up into subcomponents.
These shortcomings largely arose from (mis)applying these algorithms
outside of the domain in which they were originally developed.  The
unsatisfactory results suggested that developing a tailored cataloging
algorithm was necessary.

Sources are extracted from the Bolocam images using a seeded watershed
algorithm \citep{itk-book}.  This algorithm assigns each pixel that is
likely to contain emission to a single object based on the structure
of the emission and nearby local maxima.  Watershed algorithms are so
named since they can be used to identify the watershed associated
with a given body of water.  To make the analogy to the problem at
hand, we consider the inverted emission map of a region as a
two-dimensional surface.  Several initial markers are identified (the
seeds) associated with the lowest points of the inverted image (peaks
in the original image).  The surface is then ``flooded'' from the
starting markers at successively higher levels.  At each new level,
newly flooded regions are linked to the original seeds associating
each point on the surface with one of the starting seeds.

Source identification takes place in a two-step process. First,
significant regions of emission are identified based on the amplitude
of the signal compared to a local estimate of the noise
(\S\ref{noise}).  These sources are expanded to include adjacent,
low-significance signal (\S\ref{masking}).  Second, each region so
identified is examined and subdivided further into substructures if
there is significant contrast between the sub-objects in the region
(\S\ref{watershed_sec}).

The algorithm which accomplishes this object identification contains
user-determined parameters.  In the absence of physical priors, we
have set these parameters to reproduce a ``by-eye'' decomposition of
some of the images.  We use the same set of parameters to decompose
all images in the BGPS and we have validated the parameter selection
through in depth testing of the accuracy of the resulting catalog
(\S\ref{testing}).  Even though the choice of some parameter values
are motivated by the desired behavior of the algorithm, the influence
of these choices is well-known.

\subsection{Local Noise Estimation}
\label{noise}
\begin{figure}
\plotone{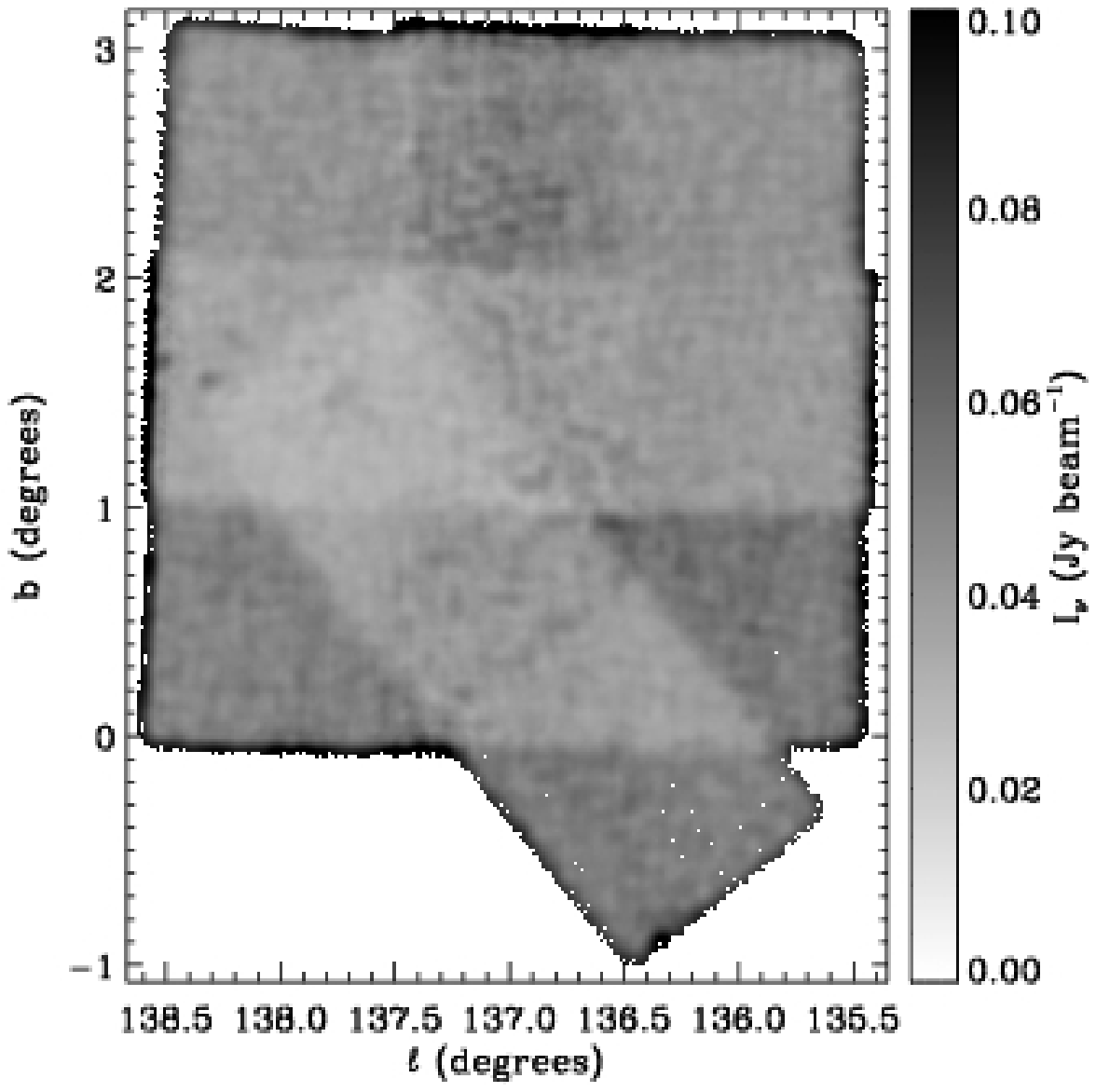}
\caption{\label{noisemap} Local variations in the noise for the W5
  region of the BGPS.  The noise map is generated from the residual
  map created in the image generation process.  The map illustrates
  variations in the noise level due to variable coverage by the survey
  and changing observing conditions.  The W5 field was selected for
  display based on its varied survey coverage resulting in the
  observable structure in the noise map.}
\end{figure}

Objects in the BGPS catalog are identified based on their significance
relative to the noise in the map.  Owing to variations in the
observing conditions and integration time across an individual BGPS
field, the underlying noise at each position can vary substantially
across a mosaic of several scan blocks.  The spatially varying noise
RMS ($\sigma$) is calculated as a function of position using the
residual maps from the image production pipeline (Paper I).  These
maps consist of the actual images less the source emission model
(where the subtraction is performed in the time-stream).  The residual
maps are used in the map reconstruction process and are nominally
source-free representations of the noise in the map.  We estimate the
RMS by calculating the median absolute deviation in an 11$\times$11
pixel box ($80''$ square) around each position in the map.  The size
of the box is chosen as a compromise between needing an adequate
sample for the estimator and the need to reproduce sharp features in
the noise map.  The median absolute deviation estimates the RMS while
being less sensitive to outliers relative to the standard deviation.
The resulting map of the RMS shows significant pixel-to-pixel
variation owing to uncertainty in the RMS estimator for finite numbers
of data.  To reduce these fluctuations, we smooth the RMS maps with a
two-dimensional Savitsky-Golay smoothing filter \citep{numrec}.  The
filter is seventh degree and smoothes over a box $12'$ on a side.  The
action of the filter is to reproduce large scale features in the map
such as the boundaries between two noisy regions while smoothing noise
regions on beam-size scales.  An example map of the RMS is shown in
Figure \ref{noisemap} after all processing (RMS estimate and
smoothing).

\begin{figure*}
\epsscale{0.8}
\plotone{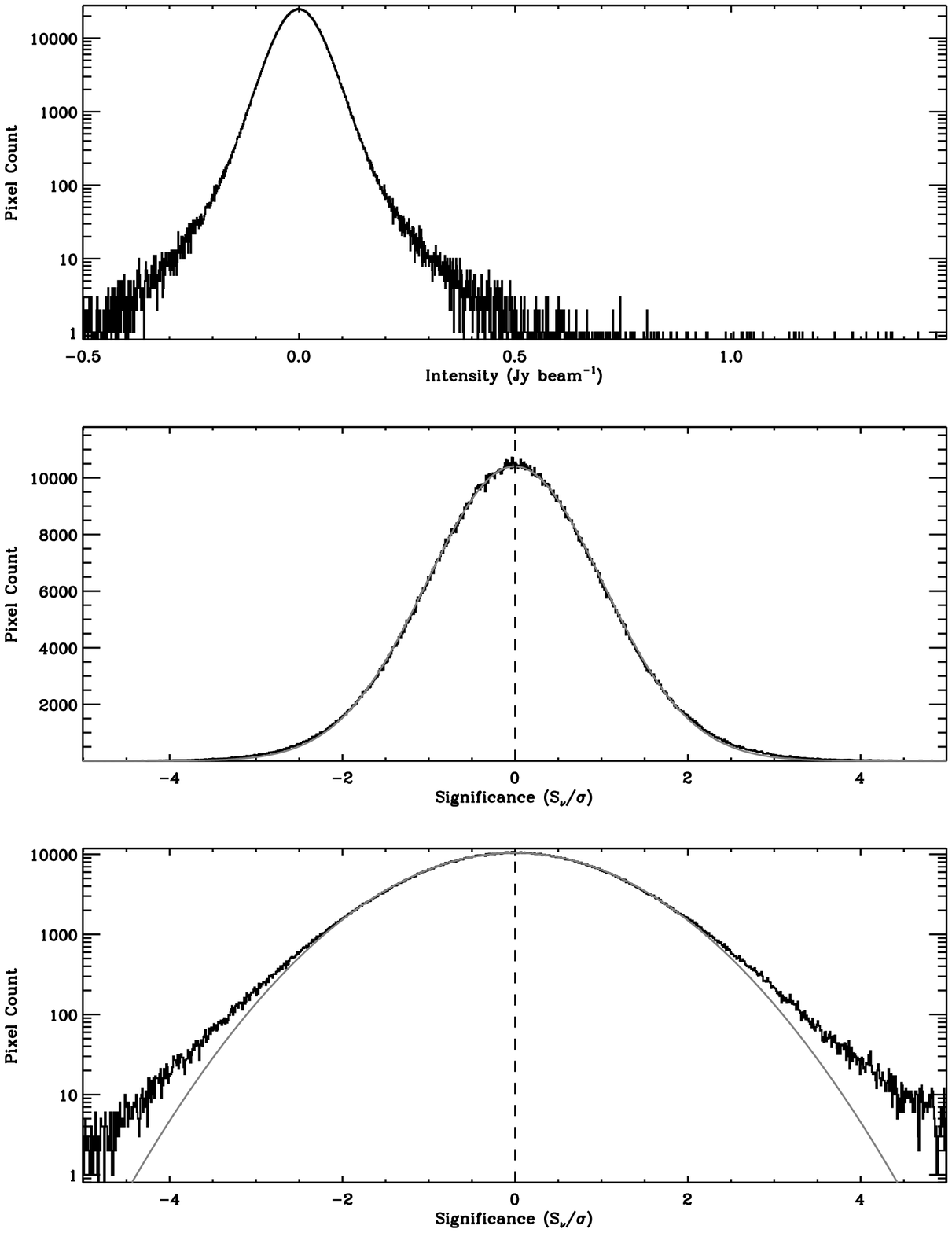}
\caption{\label{noisedist} Noise properties for the W5 BGPS field.
  The top panel shows the histogram of raw data values present in the
  image, in particular showing a large tail towards high values which
  arise because of both signal and high noise values at the map edge.
  The middle panel shows the significance distribution of the image
  after normalization by the noise estimate derived from the weight
  map.  The bottom panel repeats the data from the middle panel but
 using a logarithmic scale for the $y$-axis, highlighting the
  behavior of the distribution at large values of the significance.
  In both the middle and the bottom panel, a gray curve shows the
  Gaussian fit to the significance distribution.}  \epsscale{1.0}
\end{figure*}

The statistical noise properties of the map are shown in Figure
\ref{noisedist}, showing the intensity at each position in the map,
$I$ and the signal-to-noise ratio $I/\sigma$.  In general, generating
an error estimate from the residual map tracks the
position-to-position variation of the noise quite well.  Transforming
the data into significance space (i.e., dividing the data image by the
noise estimate) yields a normal distribution with some deviations.  As
seen in the bottom panel of the Figure in particular, there is a wing
of high significance outliers associated with the signal in the map.
There are more negative outliers than expected, and this arises from
pixel-to-pixel variations in the RMS based on local conditions.  Since
our method cannot reproduce these variations perfectly, regions with
higher true RMS than represented in the RMS map will manifest as high
significance peaks.  There are likely regions for which the RMS is an
overestimate and such regions contribute a slight excess of data near
$I/\sigma=0.0$, which is impossible to discern in the face of the
large number of data that are expected there.  We note the excess of
negative outliers may also result from incomplete removal of
astrophysical signal in the model map (see Paper I).  In principle, it
is possible to track the variations of the noise distribution on
sub-beam scales using information in the time domain.  We found,
however, that the generation of noise statistics from the time stream
is unreliable and produces non-normal distributions of significance.
However, we do use the time-stream estimate of the noise to identify
regions where the residual images are contaminated with astrophysical
signal and we replace the RMS estimate at positions with spuriously
large estimates of the noise with a local average.  A Gaussian fit to
the distribution of significance values reproduces the shape of the
distribution well with the exceptions noted above.  Fits to the normal
distribution find the central value of the distribution $\langle S
\rangle$ zero in all cases ($-0.02 \lesssim \langle I/\sigma\rangle
\lesssim 0.02$). The width of the distribution, which should be
identically 1 if $\sigma(\ell, b)$ is a perfect estimate of the noise,
is usually within 2\% of 1.00.  Hence, the significance distribution
has been transformed to the best approximation of a normal
distribution with zero mean and dispersion equal to 1.

\subsection{Identification of Significant Emission}
\label{masking}
Given an estimate of the significance of the image at every position
in the data, emission is identified as positive outlying data that are
unlikely to be generated by noise.  These data form connected {\em
  regions} of positive deviation above a given threshold.  We
initially mask the data to all regions above $2\sigma(\ell,b)$.  Owing
to image artifacts from the Bolocam mapping process, the emission
profile shows low-amplitude noise on scales smaller than a beam.  As a
consequence of this noise, holes in the mask can appear where negative
noise fluctuations drive a pixel value below the threshold.
Additionally, a set cut at a given threshold tends to produce an
artificially jagged source boundary. Finally, the PCA cleaning process
creates negative bowls around strong sources.  To reduce the influence
of these effects, we perform a morphological opening operation on the
data with a round structuring element with a diameter of a half a
beam-width \citep[see][for details]{morph}.  The opening operation
minimizes small-scale structure at the edges of the mask on scales
less than that of a beam by smoothing the edges of the mask. 

\begin{figure*}
\plotone{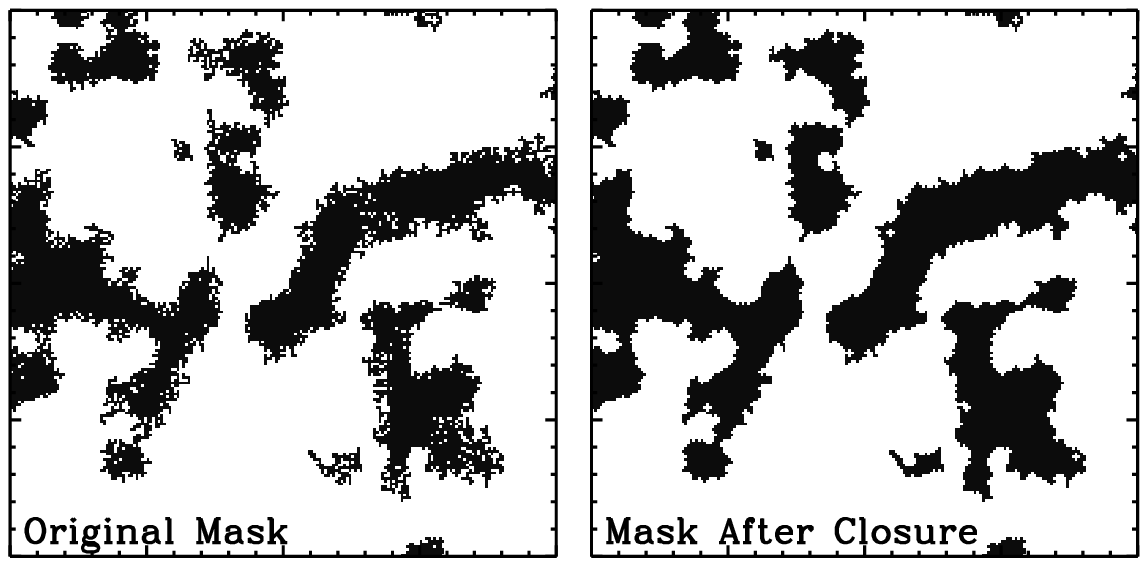}
\caption{\label{rounding}Results of applying a morphological opening
  and closing operations to a mask.  The left-hand panel shows the
  results of a clip at $2\sigma$ and the right-hand panel shows the
  mask after it has been processed by the morphological operations.
  The opening eliminates artificial structure on scales smaller than
  that of the beam.  The closure smoothes edges and eliminates holes.
  This largely cosmetic process produces smooth features which enables
  better source decomposition.  The sample region is $24'$ on a side.}
\end{figure*}

\begin{figure}
 \plotone{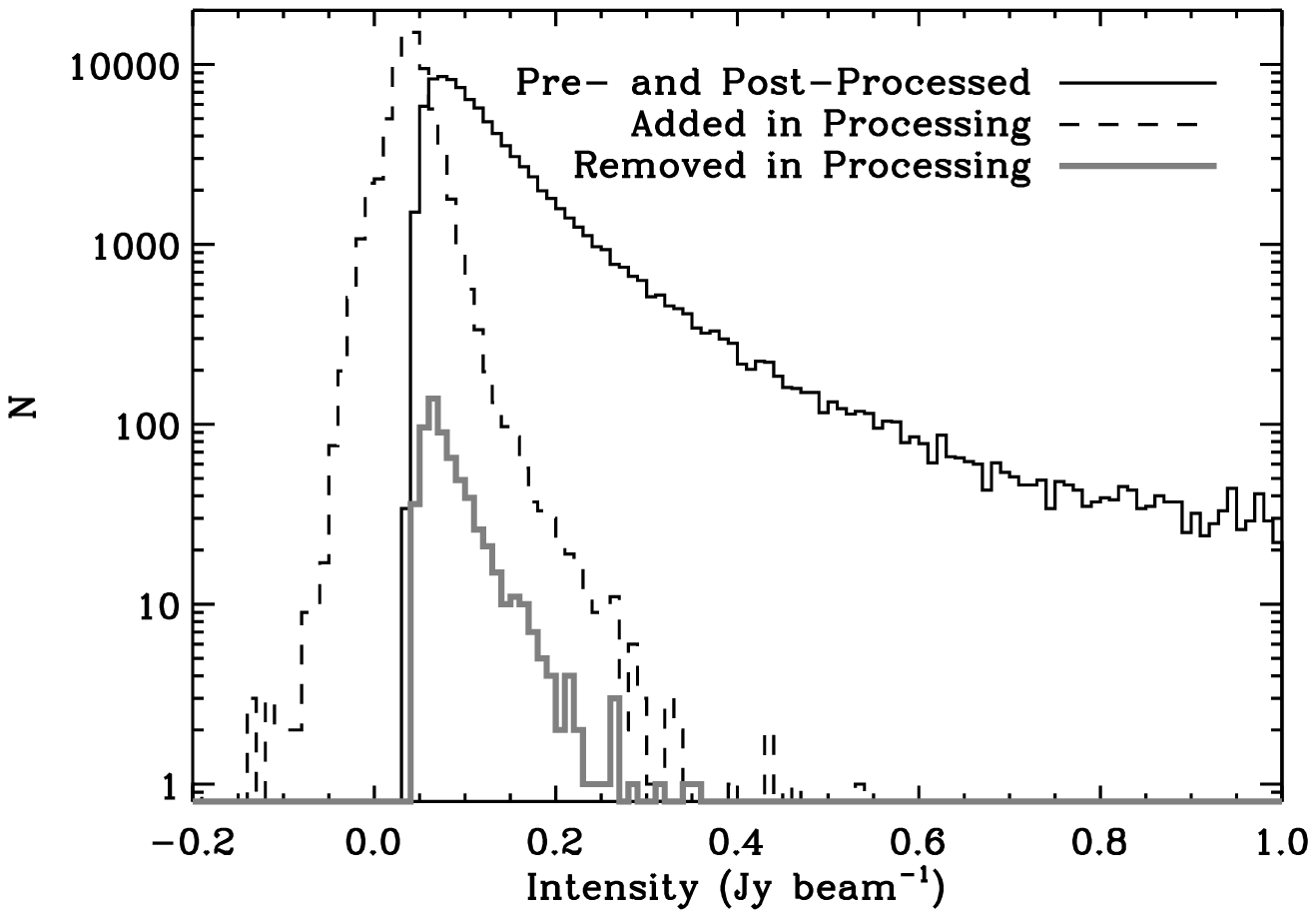}
 \caption{\label{roundingdist} The brightness distribution of pixels
   included in the emission mask pre- and post- processing by the
   opening and closing operators.  The added pixels are mostly positive
   low significance pixels (dashed line) and the removed pixels (gray
   line) are confined to relatively faint emission.  A small fraction
   (5\%) of the added pixels have negative intensities. }
 \end{figure}

Following the opening operation, the remaining regions are expanded to
include all connected regions of emission above $1\sigma(\ell,b)$
since regions of marginal significance adjacent to regions of emission
are likely to be real.  After this rejection, a morphological closing
operation is performed with the same structuring element as the
previous opening operation.  The closing operation smoothes edges and
eliminates holes in the mask.  This processing minimizes the effects
of the sharp cut made in significance on the mask, allowing
neighboring, marginal-significance data to be included in the mask.
An example of the effects for the opening and closing operations is
shown in Figure \ref{rounding}.  This results in our final mask
containing emission.  To assess the effects of the mask processing on
our data, we examined the brightness distribution of the pixels in the
mask pre- and post-processing by the opening and closing operations.
Figure \ref{roundingdist} presents the pixel brightness distributions
of the regions included in our mask both before and after applying the
operators.  The eliminated pixels are at low significance and the
added pixels include marginal emission as desired.  The processed mask
also includes some negative intensity pixels in the objects.  The
added negative pixels are only 5\% by number of the pixels added by
mask processing and are not included in the calculation of source
properties (\S\ref{properties}).

\subsection{Decomposing Regions of Emission}
\label{watershed_sec}
The emission mask is a binary array with the value 1 where the map
likely contains emission and 0 otherwise.  Identification of objects
occurs only within the emission mask region.  We consider each
spatially contiguous region of emission separately and identify
substructure within the region.  We determine whether this is
significant substructure within the map by searching for distinct
local maxima within the region.  If there is only one distinct local
maximum, the entire region is assigned to a single object.  Otherwise,
the emission is divided among the local maxima using the algorithm
described below.

\begin{figure}
\plotone{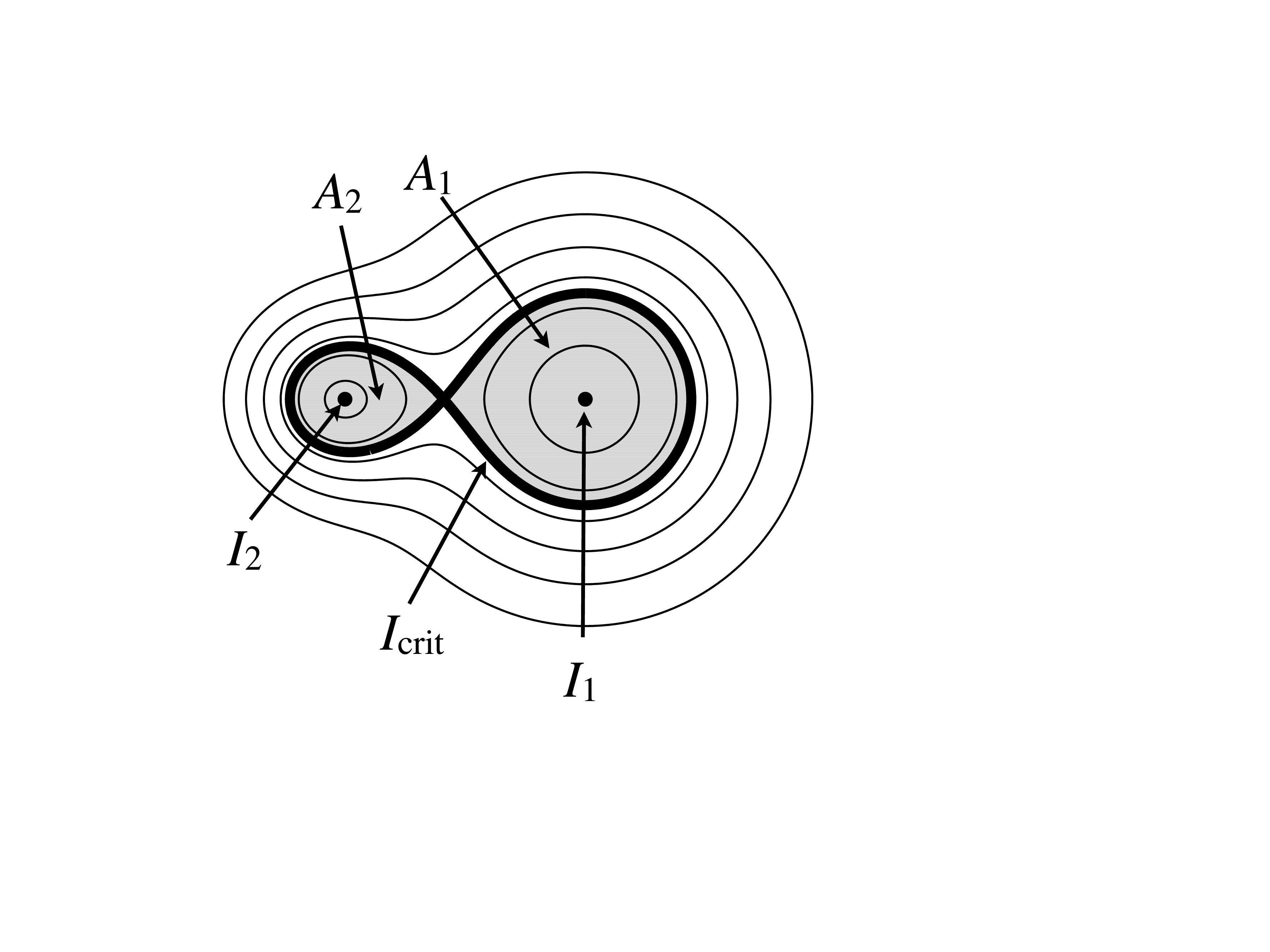}
\caption{Schematic contour diagram of elimination criterion for local
  maxima.  Each of a pair of local maximum (here $I_1$, $I_2$) has an
  area uniquely associated with that local maximum (i.e., above a
  critical contour $I_{crit}$).  If the difference between
  $I_i/\sigma-I_{crit}/\sigma$ is smaller than 0.5 then the smaller
  amplitude local maximum is removed from the list of candidate local
  maxima.  If the smaller area uniquely associated with a maximum
  (here $A_2$) is smaller than 3 pixels, then that maximum is
  removed.\label{splitschematic}}
\end{figure}

To assess whether a region has multiple distinct local maxima, we
search for all local maxima in the region requiring a local maxima to
be larger than all pixels within 1 beam FWHM of the candidate.  For
each pair of local maxima, we then identify the largest contour value
that contains both local maxima ($I_{crit}$) and measure the area
uniquely associated with each local maximum.  If either area is
smaller than 3 pixels (as it would be for a noise spike) then the
local maximum with the lowest amplitude is removed from consideration
(see Figure \ref{splitschematic}).  Furthermore, if either peak is
less than 0.5$\sigma$ above $I_{crit}$ then the maximum with the
smaller value of $I_\nu/\sigma$ is rejected.  This filter dramatically
reduces substructure in regions of emission that is due to noise
fluctuations.  We note that the filter is fairly conservative compared
to a by-eye decomposition.  As a consequence, compound objects are
de-blended less aggressively than a manual decomposition.  The
rejection criteria are evaluated for every pair of local maxima in a
connected region, so every local maximum that remains after decimation
is distinct from every other remaining maximum.

\begin{figure*}
\plottwo{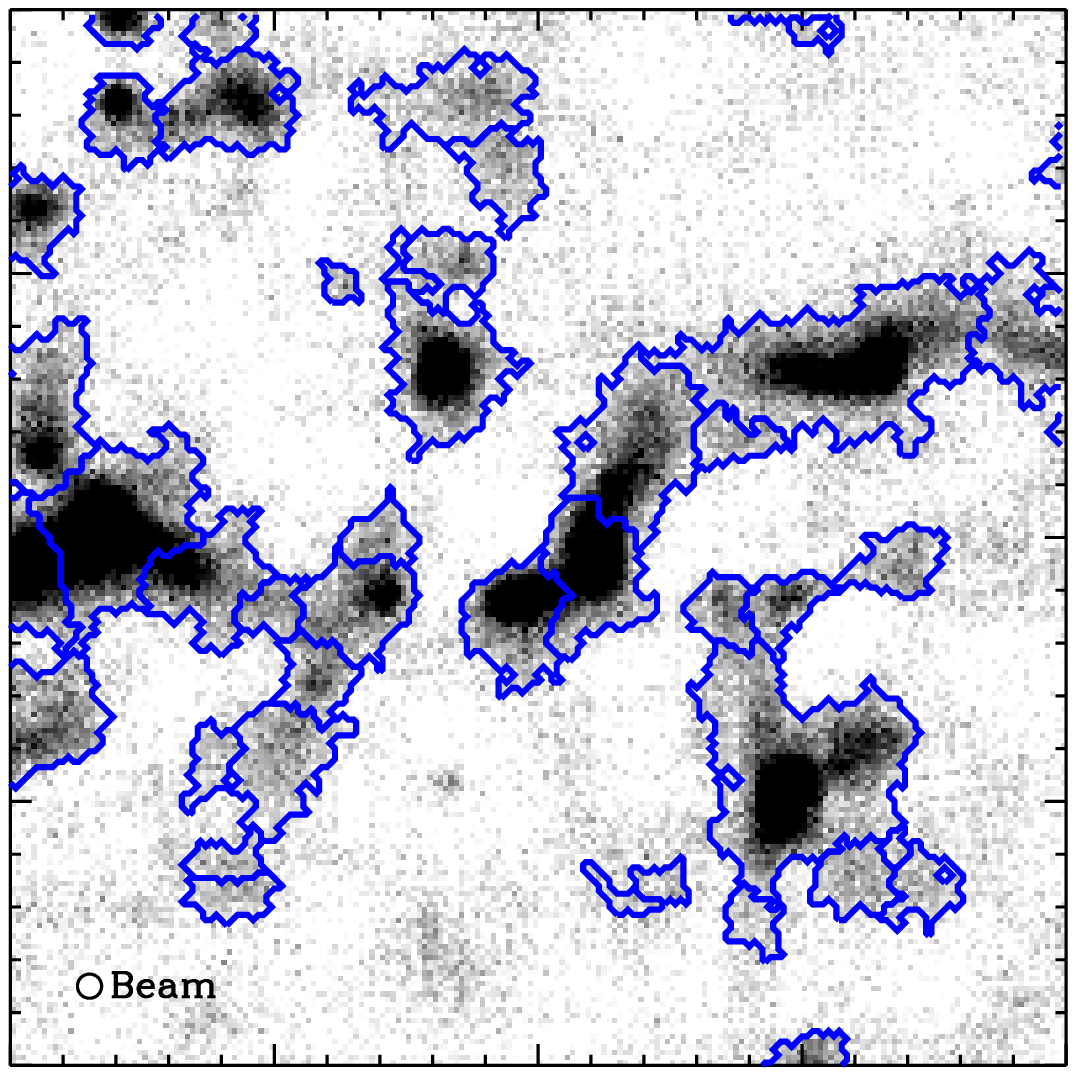}{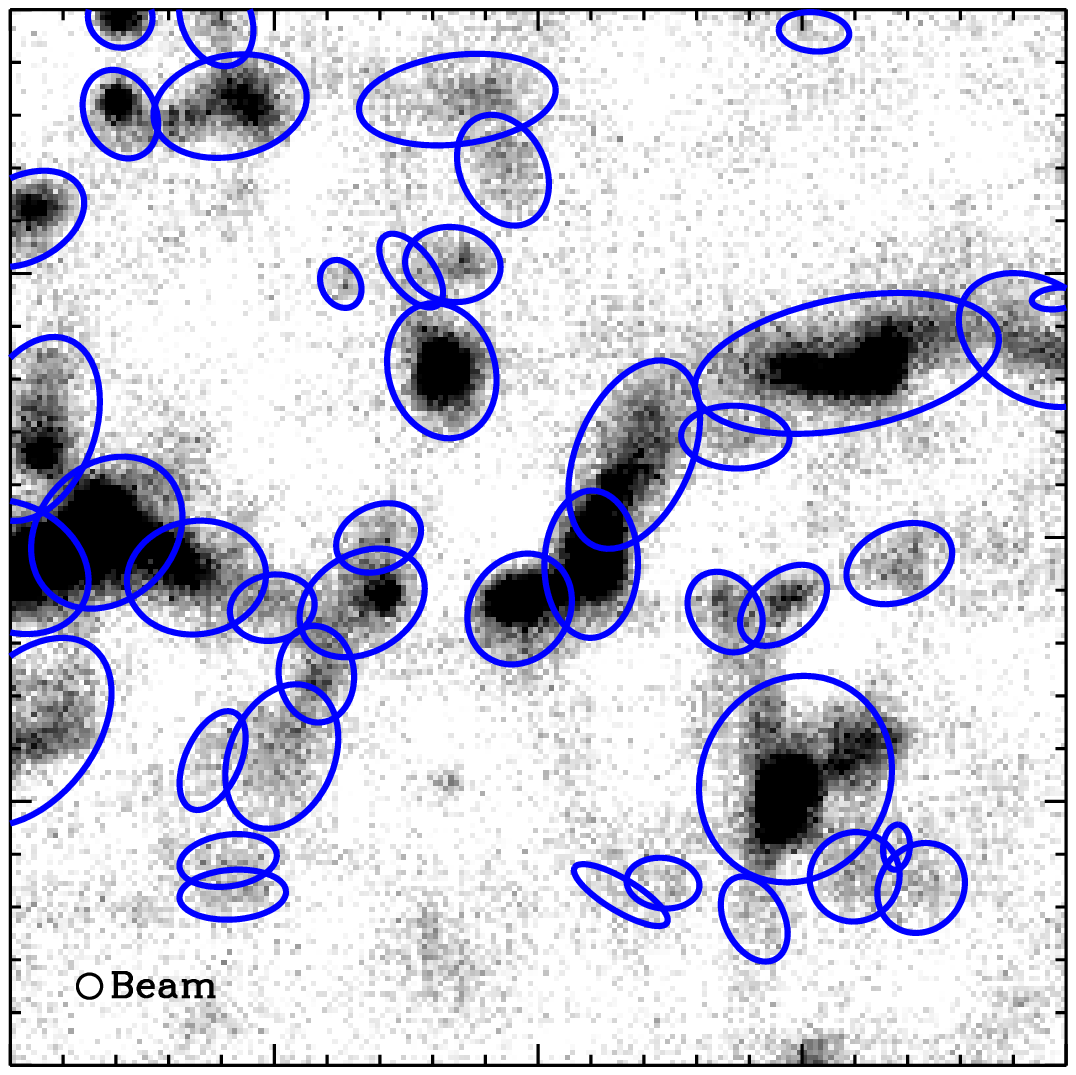}
\caption{\label{watershed} Example of the watershed algorithm applied
  to the same region shown in Figure \ref{rounding} ({\it left}) and
  the measured source shapes ({\it right}).  The grayscale image shows
  the actual BGPS map image which runs on a linear scale from 0 to 0.2
  Jy beam$^{-1}$.  In the left panel, the blue boundaries indicate the
  regions identified using the seeded watershed algorithm.  In the
  right panel, the ellipses indicate the shapes derived from the
  moment methods applied to the extracted objects.  The example region
  shown is from the original $\ell=32^{\circ}$ BGPS maps and is $24'$
  on a side.}
\end{figure*}

The final set of local maxima are thus well-separated from their
neighbors, have significant contrast with respect to the larger
complex, and have a significant portion of the image associated with
them.  Using this set of local maxima, we partition the regions of
emission into catalog objects.  This partitioning is accomplished
using a seeded watershed algorithm.  Such an algorithm is, in
principle, similar to the methods used in Clumpfind \citep{clumpfind}
in molecular line astronomy or Source EXtractor \citep{sextractor} in
optical astronomy.  To effect the decomposition, the emission is
contoured with a large number of levels (500 in this case).  For each
level beginning with the highest, the algorithm clips the emission at
that level and examines all the resulting isolated sub-regions.  For
any pixels not considered at a higher level, the algorithm assigns
those pixels to the local maximum to which they are connected by the
shortest path contained entirely within the subregion.  If there is
only one local maximum within the sub-region, the assignment of the
newly examined pixels is trivial.  The test contour levels are marched
progressively lower until all the pixels in the region are assigned to
a local maximum.  An example of the partitioning of a region is shown
in Figure \ref{watershed}.

\subsection{Measuring Source Properties}
\label{properties}
A BGPS catalog object is defined by the assignment algorithm
established above which yields a set of pixels with known Galactic
coordinates $(\ell_j,b_j)$ and associated intensities $I_j$ (Jy
beam$^{-1}$).  Source properties are determined by emission-weighted
moments over the coordinate axes of the BGPS image.  Positions in the
mask with negative intensity are not included in the calculation of
properties.  For example, the centroid Galactic longitude of an object
is
\begin{equation}
\ell_{cen} = \frac{\sum_j \ell_j I_j}{\sum_j I_j}
\end{equation}
with $b_{cen}$ defined in an analogous fashion.  The sizes of the
objects are determined from the second moments of the emission along
the coordinate axes:
\begin{eqnarray}
\sigma_\ell^2 &=& \frac{\sum_j \left(\ell_j-\ell_{cen}\right)^2
  I_j}{\sum_j I_j}\\
\sigma_b^2 &=& \frac{\sum_j \left(b_j-b_{cen}\right)^2
  I_j}{\sum_j I_j}\\
\sigma_{\ell b} &=& \frac{\sum_j
  \left(\ell_j-\ell_{cen}\right)\left(b_j-b_{cen}\right) I_j}{\sum_j
  I_j}
\end{eqnarray}
The principal axes of the flux density distribution are determined by
diagonalizing the tensor
\begin{equation}
\mathbf{I}=\left[\begin{array}{cc} \sigma_\ell^2 & \sigma_{\ell b}
\\ \sigma_{\ell b} & \sigma_b^2 \end{array}\right].
\end{equation}
The diagonalization yields the major ($\sigma_{maj}$) and minor
($\sigma_{min}$) axis dispersions as well as the position angle of the
source. Since the moments are calculated for and emission mask that
has been clipped at a positive significance level, these estimates
will slightly underestimate the sizes of the sources leading to size
estimates that could be smaller than the beam size.  Once projected
into the principal axes of the intensity distribution, the angular
radius of the object can be calculated as the geometric mean of the
deconvolved major and minor axes:
\begin{equation}
\label{radius}
\theta_R= \eta \left[(\sigma_{maj}^2-\sigma_{bm}^2)(\sigma_{min}^2-\sigma_{bm}^2)\right]^{1/4}
\end{equation}
Here $\sigma_{bm}$ is the rms size of the beam
($=\theta_{\mathrm{FWHM}}/\sqrt{8\ln 2}=14''$) and $\eta$ is a factor
that relates the rms size of the emission distribution to the angular
radius of the object determined.  The appropriate value of $\eta$ to
be used depends on the true emission distribution of the object and
its size relative to the beam.  We have elected to use a value of
$\eta=2.4$ which is the median value derived for a range of models
consisting of a spherical, emissivity distribution $\j(r)\propto
r^{\gamma}$ with $-2 < \gamma < 0$, a range of radii relative to the
beam $0.1<\theta_R/\theta_{\mathrm{FWHM}}<3$ at a range of
significance (peak signal-to-noise ranging from 3 to 100).  The range
of $\eta$ for the simulations is large, varying by more than a factor
of two.  If a specific source model is required for a follow-up study,
the appropriate value of $\eta$ should be calculated for this model
and the catalog values can be rescaled to match.

Uncertainties in the derived quantities are calculated by propagating
the pixel-wise estimates of the flux density uncertainties through the
definitions of the quantities. The coordinate axes are assumed to have
negligible errors.  The moment descriptions of position, size and
orientation are exact for Gaussian emission profiles and generate good
Gaussian approximations for more complex profiles.  An example of the
derived shapes of the catalog objects is shown in Figure
\ref{watershed} for comparison with the shapes of the regions from
which they are derived.  In comparing the images, it is clear that the
irregularly shaped catalog objects can produce elliptical sources that
vary from what a Gaussian fit would produce.  However, in most cases
the moment approximations characterize the object shapes well.

Since the extracted objects are often asymmetric, the centroid of the
emission does not correspond to the peak of emission in an object.
Because of this possible offset, we also report the coordinates of the
maximum of emission.  To minimize the effect of noise on a sub-beam
scale, we replace the map values with the median value in a $5\times
5$ pixel ($36''$) box around the position and determine the position
of the maximum value in the smoothed map.  While this reduces the
positional accuracy of the derived maximum, it dramatically improves
the stability of the estimate.  Since the peaks of emission are rarely
point sources, the loss of resolution is made up for by the increased
signal-to-noise in the map.  These coordinates are reported as
$\ell_{max}$ and $b_{max}$.  If the catalog is being used to follow-up
on bright millimeter sources in the plane these coordinates are
preferred over the centroid coordinates.  Moreover, since the maxima
are less subject to decomposition ambiguities, we name our sources
based on these coordinates: G$\ell\ell\ell$.$\ell\ell\ell\pm bb.bbb$.
The smoothed map is only used for the calculation of the position for
the maximum; all other properties are calculated from the original
BGPS map.

The flux densities of the extracted objects are determined by
measuring the flux density in an aperture and by integrating the flux
density associated with the region defined by each object in the
watershed extraction.  In the catalog, three flux density values for
each source are reported, corresponding to summing the intensity in
apertures with diameters $d_{app}=40'', 80''$, and~$120''$.  We refer
to these flux densities as $S_{40}, S_{80}$, and $S_{120}$
respectively.  Because many of the objects are identified in crowded
fields, we refrain from a general ``sky'' correction since inspection
of the noise statistics suggests that the average value of the sky is
quite close to zero.  One notable exception to this case is where
there are pronounced negative bowls in the emission around the object.
Because large apertures may contain parts of these negative bowls, the
aperture flux densities of bright objects are likely underestimated.
In addition, we also report the object flux density value ($S$), which
is determined by integrating the flux density from all pixels
contributing to the object in the watershed decomposition of the
emission:
\begin{equation}
\label{fluxobj}
S\equiv \sum_i I_i~\Delta \Omega_i,
\end{equation}
where $\Delta \Omega_i$ is the angular size of a pixel.  The object
flux density cannot, by definition of the regions, contain negative
bowls, but the bowls indicate that emission is not fully restored to
these values, which are thus likely underestimates of the true flux
density for bright objects.  Since objects typically have angular
radii $\sim 60''$, $120''$ apertures do not contain all the flux
density.  Hence, $S\gtrsim S_{120}> S_{80} > S_{40}$.  Since the
surface brightness at each position in the map has an associated
uncertainty (\S\ref{noise}), we calculate the the uncertainty in the
flux density measurements by summing the uncertainties in quadrature.
We further account for the uncertainty in the size of the beam owing
to the uncertainties of the bolometer positions in the focal plane of
the instrument (see Paper I).  The latter effect produces a fractional
uncertainty in the beam area of $\delta
\Omega_{bm}/\Omega_{bm}=0.06$.  Thus, the uncertainty in a flux
density measurement (either aperture or integrated) is:
\begin{equation}
\delta S \equiv \sqrt{\frac{\Omega_i}{\Omega_{bm}}\sum_i
  \sigma^2_i+\left(\frac{\delta \Omega_{bm}}{\Omega_{bm}}S\right)^2}
\label{fluxerror}
\end{equation}
where the ratio $\Omega_i/\Omega_{bm}$ accounts for the multiple
pixels per independent resolution element.  The flux density errors do
{\em not} account for any systematic uncertainty in the calibration of
the image data which is estimated to be less than 10\% (Paper I).  We
discuss the flux densities in more detail in Section \ref{fluxes}
below.

\section{Source Recovery Experiments}
\label{testing}

In order to assess the properties of catalog sources compared to the
actual distribution of emission on the sky, we have created simulated
maps of the emission, inserting sources with known properties.  The
generation of these simulated images relies on the actual observed
data for the $\ell=111^{\circ}$ region of the survey.  This field was
selected for analysis based on its uniform noise properties and
well-recovered astrophysical signal.  Most of the astrophysical signal
is identified in the time series data from the instrument and removed,
leaving only irreducible (i.e., shot) noise in the time series with
small contributions due to atmospheric, instrumental (e.g., the pickup
of vibrations) and astrophysical signals (see Paper I for more
detail).  Then a series of Gaussian objects was sampled into the data
time stream.  The simulation time streams are then processed in a
fashion identical to the final BGPS survey images resulting in a suite
of test images to test the catalog algorithm on.  Gaussian brightness
profiles do not adequately represent all the structure seen in the
BGPS maps, but they do capture the essential features of compact
sources.  The complicated interplay of extended structure with the
mapping process likely affects the validity of the following tests.
However, the behavior for point-like or small sources should be well
reproduced.  Finally, some of the astrophysical signal remains in the
residual image which affects some of our tests to a small degree.

\subsection{Completeness}
\begin{figure}
\plotone{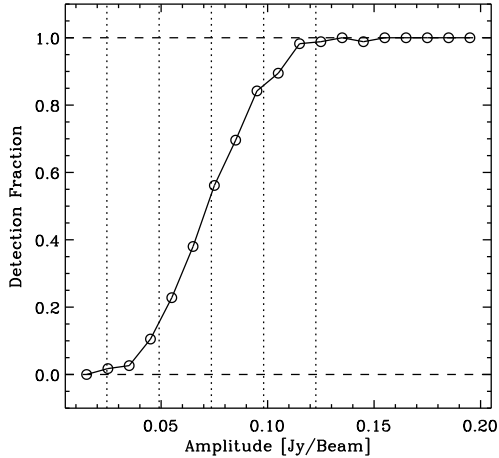}
\caption{\label{complim} Completeness fraction derived from fake
  source tests in a survey field.  The fraction of sources recovered
  is plotted as a function of input source flux density.  The vertical dotted
  lines indicate \{1,2,3,4,5\}$\sigma$.  The completeness tests
  indicate that the catalog is complete at the $>99\%$ limit for
  source flux densities  $\ge 5\sigma$.}
\end{figure}

The primary test of the algorithm is the degree of completeness of the
catalog at various flux density limits.  To this end, we have inserted
a set of point sources into the time stream with flux densities
uniformly distributed between 0.01 Jy to 0.2 Jy arranged in a grid.
We then process the simulated data with the source extraction
algorithm and compare the input and output source catalog.  The
results of the completeness study are shown in Figure \ref{complim}.
For this field, the mean noise level is $\sigma=24$~mJy~beam$^{-1}$.
The catalog is $>99\%$ complete at the 5$\sigma$ level.  Below this
level, the completeness fraction rolls off until no sources are
detected at 2$\sigma$.  For this regular distribution of sources,
there are no false detections throughout the catalog (even at the
2$\sigma$ level).  The catalog parameters described above are selected
to minimize the number of false detections.  Even though the
thresholding of the map is performed at the relatively low 2.0$\sigma$
level, the wealth of restrictions placed on the extracted regions
combine to raise the effective completeness limit to $5\sigma$.  We
note that the grid arrangement of test sources means that confusion
effects are not accounted for in the flux limit.

\begin{figure*}
\plotone{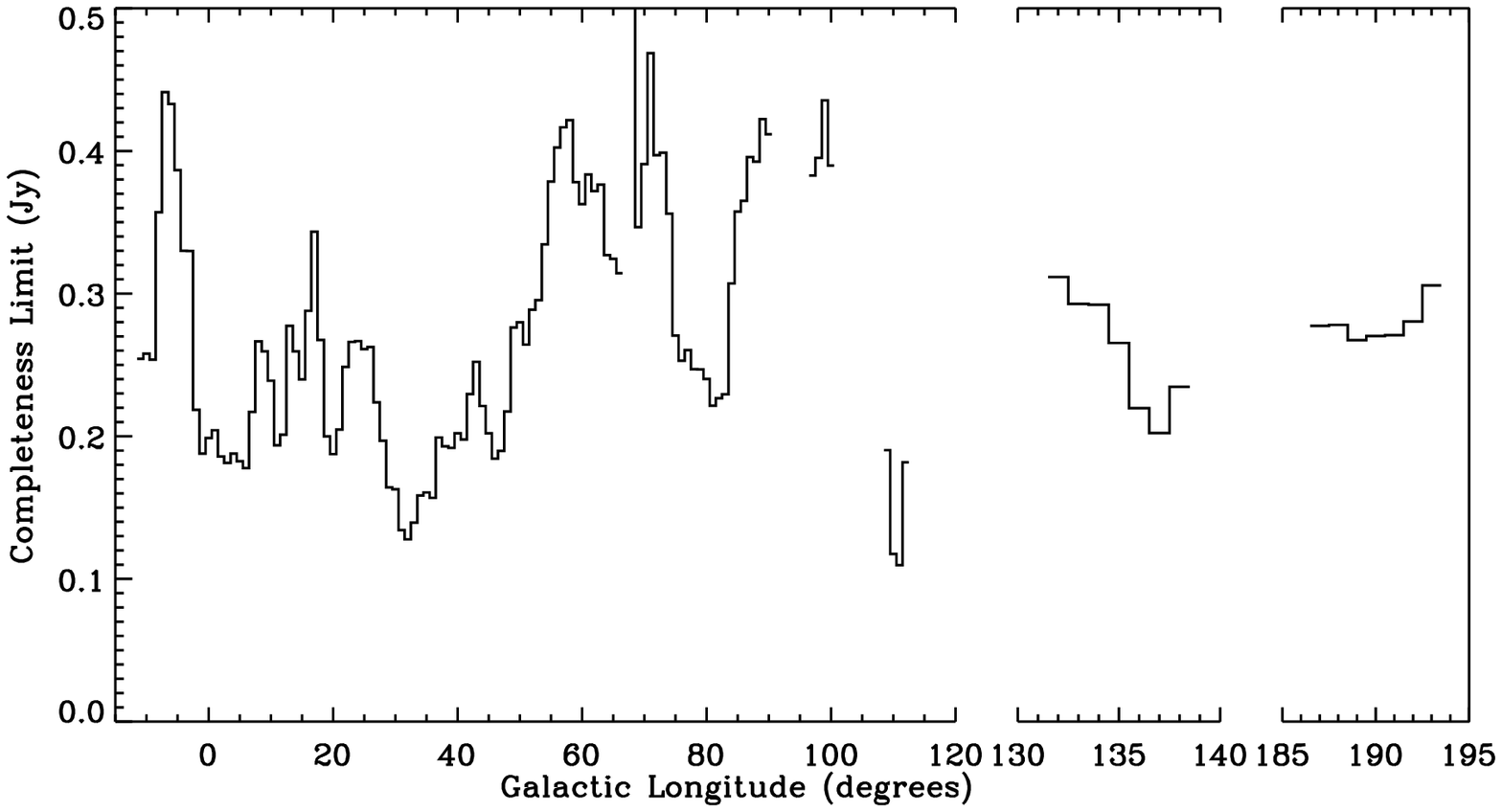}
\caption{\label{complimlon} The 99\% completeness limit in flux
  density for point sources plotted as a function of Galactic
  longitude.  The completeness limit is 5 times the median RMS value
  in $1^{\circ}$ bins across the survey.  The significant variation in
  the completeness limit stems from the variable observing conditions
  and integration time for each field in the survey.} \epsscale{1.0}
\end{figure*}

In Figure \ref{complimlon}, we present the estimated completeness
limit as a function of Galactic longitude for the BGPS.  The
$\ell=111^{\circ}$ region has a relatively low value for the noise RMS
with several fields being up to a factor of 4 larger than the
completeness test field.  We assume that the completeness limit scales
with the local RMS.  The assumption is motivated since all catalog
source extraction and decomposition is performed in the significance
map rather than in the image units.  Thus, the absolute value of the
image data should not matter.  Over $>98\%$ of the cataloged area, the
completeness limit is everywhere less than 0.4 Jy, thus the survey, as
a whole can be taken to be 98\% complete at the 0.4 Jy level.

\subsection{Decomposition Fidelity}

\begin{figure*}
\plottwo{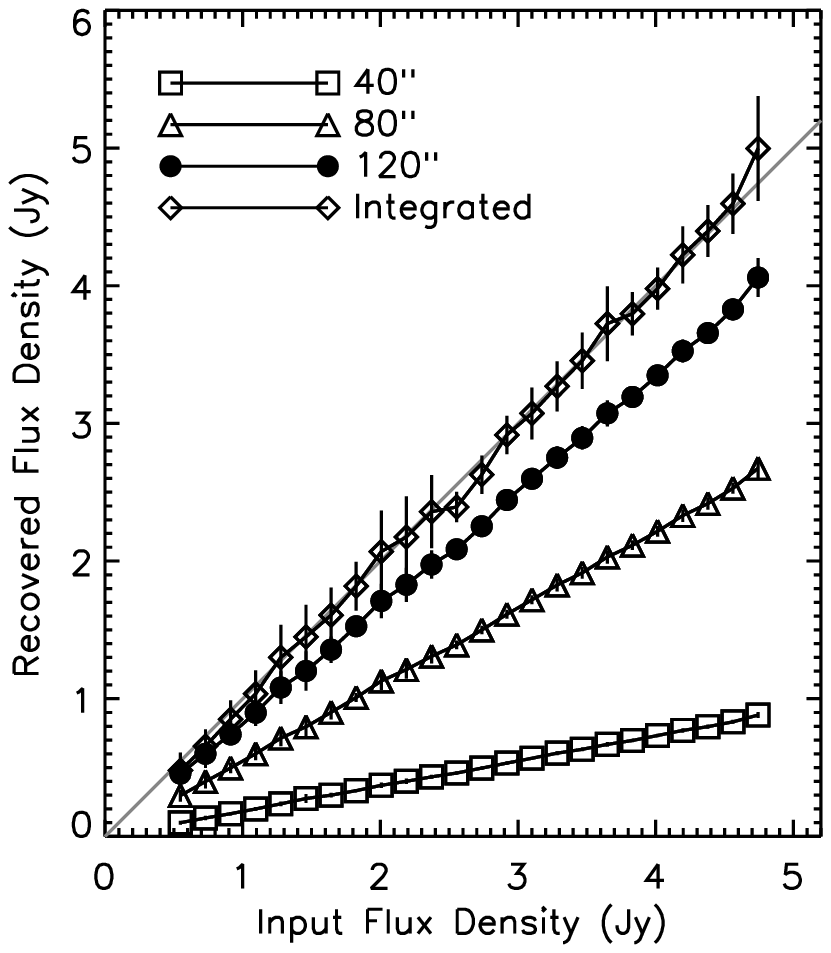}{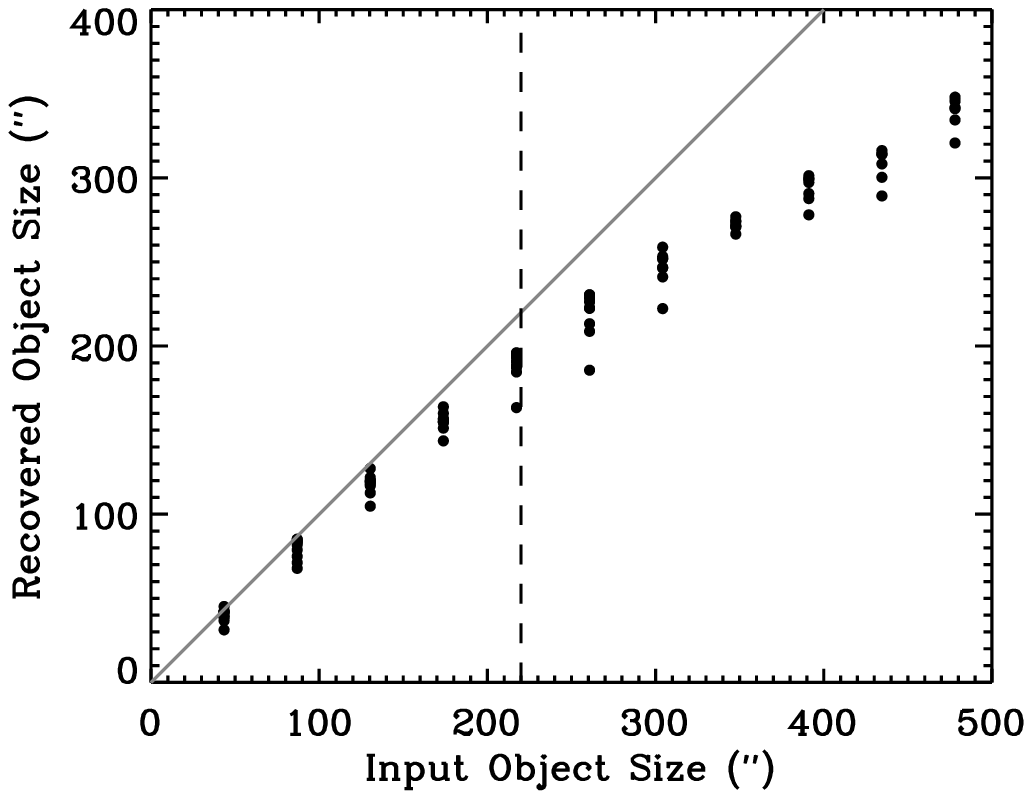}
\caption{\label{fluxrad} Recovery of source flux density and source
  size for simulated observations.  The left-hand panel compares input
  and recovered source flux densities for resolved objects (FWHM =
  2.4$\theta_{beam}=79''$).  The right-hand panel shows the size
  measured in the simulated objects for an array of input object
  sizes.  The vertical dashed line at $210''$ indicates the largest
  major-axis size of objects actually found in the BGPS implying
  source sizes are typically well recovered. }
\end{figure*}

In addition to completeness limits, we investigated the fidelity of
the algorithm by comparing extracted source properties to the input
source distributions.  We compared the extracted flux density values
to the input flux density values as shown in Figure \ref{fluxrad}.
These simulations compare input and recovered properties for resolved
sources with intrinsic FWHMs equal to 2.4 times that of the beam.  The
flux densities of the sources are varied over a significant range, and
we measure the recovered values for the flux density using the 40$''$,
80$''$ and $120''$ apertures and we also compute the integrated flux
density in the object.  Since the small apertures only contain a
fraction of the total flux density, their average flux density
recovery is naturally smaller than the input.  As the aperture becomes
better matched to the source size, the flux density recovery improves,
in agreement with analytic expectations based on the source profile.
Since many objects are larger than the $40''$ and $80''$ apertures,
catalog flux density values can be viewed in terms of surface
brightnesses within these apertures.

We completed an additional suite of simulations, including sources
with sizes significantly larger than the beam.  These simulations are
used in Paper I to test the effect of the deconvolution process, but
we also examine the results of the source extraction.  In Figure
\ref{fluxrad} we show the recovered source size as a function of input
size.  In general, smaller sources are better recovered with objects
much larger than the beam having a smaller size found.  Radii of
objects become biased for sources with sizes $\gtrsim 200''$,
consistent with the experiments on deconvolution discussed in Paper I.
Because of the image processing methods used, objects with sizes
larger than this should be treated with caution.  However, we note
that no sources in the actual catalog are found with sizes $>220''$
and the mean size is only $60''$ so objects of these sizes and their
properties are likely to be recovered.  The catalog algorithm thus
does not introduce any {\it additional} bias into the properties of
the recovered sources.  However, the sizes of objects may not be
completely recovered in the mapping process and thus the extracted
properties may not accurately reflect the true properties of sources
because of limitations in the map making process. These map
properties should be viewed with the caveats presented in Paper I.

\subsection{Deblending}

\begin{figure}
\plotone{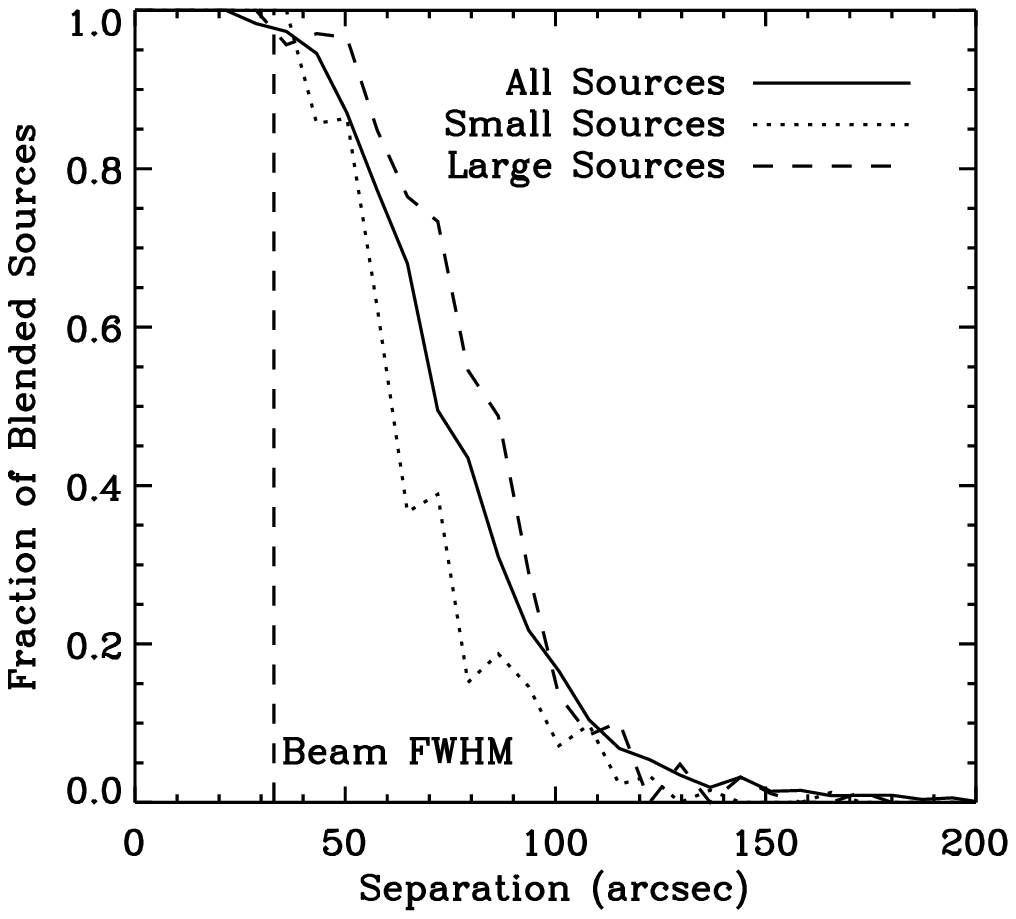}
\caption{\label{deblending} Fraction of source pairs that remain
  blended as a function of source pair separation. The solid line
  shows the blending fraction for all source pairs.  Since the sources
  are elliptical Gaussians, the dashed and dotted curves show the
  same function plotted only for the source pairs with mean sizes in
  the top and bottom 15\%, indicating that large resolved sources are
  more difficult to separate algorithmically.}
\end{figure}

A primary question of the catalog algorithm is how well it separates
individual sources into their appropriate subcomponents.  Any approach
to source extraction and decomposition will be unable to separate some
source pairs accurately.  However, the limiting separation for source
extraction is determined by the method.  Our approach is further
complicated by the assignment of emission from every position in the
catalog to at most one object.  Thus, blended sources are difficult to
extract separately unless the individual local maxima that define the
sources are apparent according to the criteria in
\S\ref{watershed_sec}.  To assess how the catalog algorithm recovers
sources, we generated maps of sources with a random distribution in
position and processed the maps with the catalog algorithm.  The input
sources have elliptical Gaussian profiles with a uniform distribution
of major and minor axis sizes between our beam size ($33''$) and
$72''$, which represent typical objects seen in the BGPS maps.  The
input sources are oriented randomly.  For each pair of input sources,
we determine whether the pair is assigned to the same catalog object.
We then measure the fraction of blended sources as a function of
separation.  The results of the experiment are shown in Figure
\ref{deblending} which indicates that the algorithm resolves most
sources that are separated by $>75''$ or 2 beam FWHM.  For point
sources, the nominal resolution limit is $\approx 1$ beam FWHM;
however the algorithm requires $\sim$2 FWHM between local maxima so
this limit on deblending is consistent with the design.  Furthermore,
the test objects are resolved, which complicates the analysis.  Thus,
Figure \ref{deblending} also indicates the fraction of blended sources
for pairs of the smallest sources and for pairs of the largest sources
(i.e., both sources are in the bottom 15\% of the size distribution or
the top 15\% respectively).  In separating the data based on size, we
can assess the degree to which source structure will complicate
deblending.  As expected, pairs of small sources are easier to resolve
(50\% are separated for distances of $60''$) and pairs of large
sources more difficult to deblend (50\% at $85''$).  The large source
curve crosses the population curve at large separations ($120''$)
because of small number statistics rather than any algorithmic
feature.

\section{The Catalog}
\label{thecatalog}
The BGPS catalog contains 8358 sources extracted using the algorithm
described in \S\ref{howto}.  An excerpt from the catalog is shown in
Table \ref{cattable}.  The full catalog and the BGPS image data are
hosted at the Infrared Processing and Analysis
Center\footnote{\url{http://irsa.ipac.caltech.edu/data/BOLOCAM\_GPS/}}.  

\begin{deluxetable*}{rlrrrrrrrrcccc}
\small
\tabletypesize{\scriptsize}
\tablewidth{0pt}
\tablecaption{The BGPS Catalog (Excerpt)\tablenotemark{a} \label{cattable}}
\tablehead{\colhead{No.} & \colhead{Name} 
& \colhead{$\ell_{\mathrm{max}}$} & \colhead{$b_{\mathrm{max}}$}
& \colhead{$\ell$} & \colhead{$b$} &
  \colhead{$\sigma_{maj}$} & \colhead{$\sigma_{min}$} & \colhead{PA} &
  \colhead{$\theta_R$} & \colhead{$S_{40}$} &\colhead{$S_{80}$} &
  \colhead{$S_{120}$}  & \colhead{$S$} \\
\colhead{} & \colhead{} & \colhead{($^{\circ}$)} &
\colhead{($^{\circ}$)} & \colhead{($^{\circ}$)} &
\colhead{($^{\circ}$)} & \colhead{($''$)}
& \colhead{($''$)} & \colhead{($^{\circ}$)}& \colhead{($''$)}
& \colhead{(Jy)} & \colhead{(Jy)} & 
\colhead{(Jy)} & \colhead{(Jy)}\\
\colhead{(1)} & \colhead{(2)} & \colhead{(3)} & \colhead{(4)} &
\colhead{(5)} & \colhead{(6)} & \colhead{(7)} & \colhead{(8)} &
\colhead{(9)} & \colhead{(10)} & \colhead{(11)} & \colhead{(12)} &
\colhead{(13)} & \colhead{(14)}}
\startdata
1 & $\mathrm{G000.000+00.057}$ & 0.000 & $0.057$ & 0.004 & $0.063$ & 32 & 17 & 89 & 38 & 0.19$\pm$0.06 & 0.56$\pm$0.11 & 1.01$\pm$0.18 & 0.67$\pm$0.12 \\ 
2 & $\mathrm{G000.004+00.277}$ & 0.004 & $0.277$ & 0.007 & $0.276$ & 39 & 34 & 145 & 80 & 0.21$\pm$0.04 & 0.59$\pm$0.09 & 0.99$\pm$0.13 & 1.40$\pm$0.17 \\ 
3 & $\mathrm{G000.006-00.135}$ & 0.006 & $-0.135$ & 0.019 & $-0.138$ & 43 & 32 & 115 & 82 & 0.22$\pm$0.05 & 0.71$\pm$0.11 & 1.39$\pm$0.17 & 1.41$\pm$0.19 \\ 
4 & $\mathrm{G000.010+00.157}$ & 0.010 & $0.157$ & 0.019 & $0.156$ & 62 & 24 & 80 & 81 & 0.61$\pm$0.06 & 1.62$\pm$0.15 & 2.33$\pm$0.21 & 3.55$\pm$0.30 \\ 
5 & $\mathrm{G000.016-00.017}$ & 0.016 & $-0.017$ & 0.012 & $-0.014$ & 50 & 35 & 44 & 93 & 1.24$\pm$0.10 & 3.68$\pm$0.26 & 6.23$\pm$0.43 & 10.94$\pm$0.74 \\ 
6 & $\mathrm{G000.018-00.431}$ & 0.018 & $-0.431$ & 0.016 & $-0.431$ & 23 & 8 & 97 & \nodata & 0.08$\pm$0.05 & 0.16$\pm$0.09 & 0.19$\pm$0.14 & 0.15$\pm$0.07 \\ 
7 & $\mathrm{G000.020+00.033}$ & 0.020 & $0.033$ & 0.019 & $0.036$ & 38 & 32 & 54 & 77 & 0.83$\pm$0.09 & 2.68$\pm$0.22 & 4.86$\pm$0.37 & 6.07$\pm$0.46 \\ 
8 & $\mathrm{G000.020-00.051}$ & 0.020 & $-0.051$ & 0.021 & $-0.052$ & 34 & 26 & 85 & 62 & 1.14$\pm$0.10 & 3.43$\pm$0.25 & 5.75$\pm$0.41 & 6.37$\pm$0.46 \\ 
9 & $\mathrm{G000.022+00.251}$ & 0.022 & $0.251$ & 0.025 & $0.250$ & 26 & 14 & 90 & \nodata & 0.10$\pm$0.04 & 0.24$\pm$0.09 & 0.43$\pm$0.13 & 0.26$\pm$0.08 \\ 
10 & $\mathrm{G000.034-00.437}$ & 0.034 & $-0.437$ & 0.037 & $-0.438$ & 17 & 11 & 92 & \nodata & 0.06$\pm$0.05 & 0.14$\pm$0.09 & 0.18$\pm$0.14 & 0.11$\pm$0.06 \\ 
11 & $\mathrm{G000.042+00.205}$ & 0.042 & $0.205$ & 0.041 & $0.208$ & 44 & 24 & 155 & 70 & 0.36$\pm$0.05 & 1.21$\pm$0.12 & 2.15$\pm$0.19 & 2.61$\pm$0.23 \\ 
12 & $\mathrm{G000.044-00.133}$ & 0.044 & $-0.133$ & 0.049 & $-0.135$ & 29 & 20 & 97 & 45 & 0.14$\pm$0.05 & 0.42$\pm$0.10 & 0.76$\pm$0.16 & 0.62$\pm$0.12 \\ 
13 & $\mathrm{G000.044-00.285}$ & 0.044 & $-0.285$ & 0.051 & $-0.282$ & 26 & 10 & 137 & \nodata & 0.08$\pm$0.04 & 0.33$\pm$0.08 & 0.60$\pm$0.13 & 0.15$\pm$0.06 \\ 
14 & $\mathrm{G000.046+00.111}$ & 0.046 & $0.111$ & 0.049 & $0.108$ & 39 & 24 & 142 & 64 & 0.33$\pm$0.05 & 0.92$\pm$0.11 & 1.40$\pm$0.16 & 1.83$\pm$0.20 \\ 
15 & $\mathrm{G000.046-00.041}$ & 0.046 & $-0.041$ & 0.047 & $-0.037$ & 34 & 29 & 168 & 67 & 0.47$\pm$0.07 & 1.67$\pm$0.17 & 3.36$\pm$0.28 & 3.21$\pm$0.28 \\ 
16 & $\mathrm{G000.046-00.297}$ & 0.046 & $-0.297$ & 0.043 & $-0.294$ & 30 & 21 & 141 & 49 & 0.11$\pm$0.04 & 0.32$\pm$0.08 & 0.52$\pm$0.12 & 0.48$\pm$0.10 \\ 
17 & $\mathrm{G000.048+00.065}$ & 0.048 & $0.065$ & 0.047 & $0.070$ & 13 & 12 & 127 & \nodata & 0.14$\pm$0.05 & 0.56$\pm$0.11 & 1.34$\pm$0.17 & 0.19$\pm$0.06 \\ 
18 & $\mathrm{G000.052+00.027}$ & 0.052 & $0.027$ & 0.048 & $0.029$ & 39 & 33 & 161 & 79 & 1.30$\pm$0.11 & 3.78$\pm$0.27 & 6.05$\pm$0.43 & 8.10$\pm$0.56 \\ 
19 & $\mathrm{G000.052-00.159}$ & 0.052 & $-0.159$ & 0.051 & $-0.159$ & 21 & 19 & 48 & 33 & 0.11$\pm$0.05 & 0.33$\pm$0.09 & 0.67$\pm$0.14 & 0.33$\pm$0.09 \\ 
20 & $\mathrm{G000.054-00.209}$ & 0.054 & $-0.209$ & 0.055 & $-0.207$ & 37 & 36 & 73 & 80 & 0.87$\pm$0.07 & 2.56$\pm$0.18 & 4.09$\pm$0.29 & 5.76$\pm$0.41 \\ 
21 & $\mathrm{G000.066+00.209}$ & 0.066 & $0.209$ & 0.067 & $0.211$ & 34 & 28 & 115 & 66 & 0.31$\pm$0.04 & 1.00$\pm$0.10 & 1.80$\pm$0.16 & 1.86$\pm$0.17 \\ 
22 & $\mathrm{G000.066-00.079}$ & 0.066 & $-0.079$ & 0.068 & $-0.076$ & 41 & 31 & 106 & 79 & 2.20$\pm$0.16 & 6.92$\pm$0.45 & 11.24$\pm$0.72 & 15.98$\pm$1.03 \\ 
23 & $\mathrm{G000.068+00.241}$ & 0.068 & $0.241$ & 0.067 & $0.244$ & 26 & 17 & 52 & 37 & 0.09$\pm$0.04 & 0.25$\pm$0.08 & 0.37$\pm$0.12 & 0.34$\pm$0.09 \\ 
24 & $\mathrm{G000.070+00.175}$ & 0.070 & $0.175$ & 0.075 & $0.173$ & 47 & 38 & 159 & 94 & 0.49$\pm$0.05 & 1.52$\pm$0.12 & 2.71$\pm$0.20 & 4.86$\pm$0.34 \\ 
25 & $\mathrm{G000.070-00.037}$ & 0.070 & $-0.037$ & 0.071 & $-0.035$ & 27 & 20 & 36 & 43 & 0.29$\pm$0.06 & 0.85$\pm$0.13 & 1.58$\pm$0.21 & 1.17$\pm$0.17 \\ 

\enddata
\tablecomments{(1) Running Source Number. (2) Name derived from
  Galactic coordinates of the maximum intensity in the object. (3)-(4)
  Galactic coordinates of maximum intensity in the catalog
  object. (5)-(6) Galactic coordinates of emission centroid. (7)-(9)
  Major and minor axis $1/e$ widths and position angle of source. (10)
  Deconvolved angular size of source.  (11)-(13) Flux densities
  derived for $40''$, $80''$ and $120''$ apertures. (14) Integrated
  flux density in the object.  The derivation of these quantities is
  described in \S\ref{properties}}
\end{deluxetable*}

\section{The Physical Properties of Catalog Sources}
\label{physical}
The BGPS data contain a wealth of sources extracted from the images
with a wide variety of source properties.  The physical meaning of a
catalog source remains unclear, however.  Ascribing nomenclature such
as a ``clump'' or a ``core'' to catalog objects may be misleading in
certain cases as these terms can refer to specific density, mass and
size regimes \citep[e.g.,][]{wbm-ppiv}. In this section, we explore
the range of parameter space a BGPS catalog object may occupy.  We
also refer readers to two additional discussions of BGPS sources in
the Galactic center \citep{bgps-galcen} and in the Gem OB1
star forming region in the outer Galaxy \citep{bgps-gemob1}.
These two studies consider the source properties in regions where most
of the sources are likely found at a known distance.  In contrast,
blind observations of the Galactic plane can span a range of
distances.

To facilitate understanding BGPS source properties, we include Robert
F. Byrd Green Bank Telescope (GBT) observations of the NH$_3$(1,1) and
(2,2) inversion transitions toward a set of BGPS detections in the
$\ell\sim 31^{\circ}$ field.  In making this comparison, we assume
that the ammonia emission comes from the same regions as are
responsible for the millimeter continuum emission, an assumption
supported by studies in local molecular clouds
\citep[e.g.,][]{nh3-perseus,friesen-oph}.  These data are presented
formally and in much greater depth elsewhere \citep{bgps-gemob1}.
Here we note that the ammonia data provide complementary information
on the line-of-sight velocities ($V_{\mathrm{LSR}}$), line widths
$\sigma_V$ and kinetic temperatures of dense gas $T_K$.  The velocity
information, in particular, yields distance estimates to the sources
by assuming that the radial motion of the line emitters arises
entirely from their orbital motion around the Galactic center.  We
then use the kinematic model of \citet{reid-kindist} to estimate the
distances to BGPS objects.  The Reid et al.~model uses VLBI parallaxes
to establish the distances to several high mass star forming
regions. One of the primary results of their work is a differentiation
between the kinematic properties of high mass star forming regions and
other components of the Galactic disk.  Since the BGPS sources are
likely associated with (high-mass) star forming regions, their refined
model is appropriate for our work.

The ammonia observations were taken before the BGPS catalog was
complete, and the GBT targets were selected by-eye from an early
version of the BGPS images.  Hence, there is not a one-to-one
correspondence between BGPS catalog sources and ammonia observations.
We match each ammonia spectrum to the BGPS catalog source for which
the pointing center for the GBT observation falls into the region
assigned to the BGPS source.  If multiple GBT observations fall within
a BGPS source, we consider only the spectrum with the highest
integrated intensity in the (1,1) transition.  In this fashion, we
match 42 ammonia spectra with BGPS sources.  We estimate the line
properties of the ammonia spectra (i.e., $V_{LSR}$, $\sigma_V$, $T_K$)
using the model of \citet{nh3-perseus}.  Of particular note, there is
only one velocity component of ammonia emission along each line of
sight examined, supporting our association between BGPS emission and
ammonia emission for these sources. 

\begin{figure}
\plotone{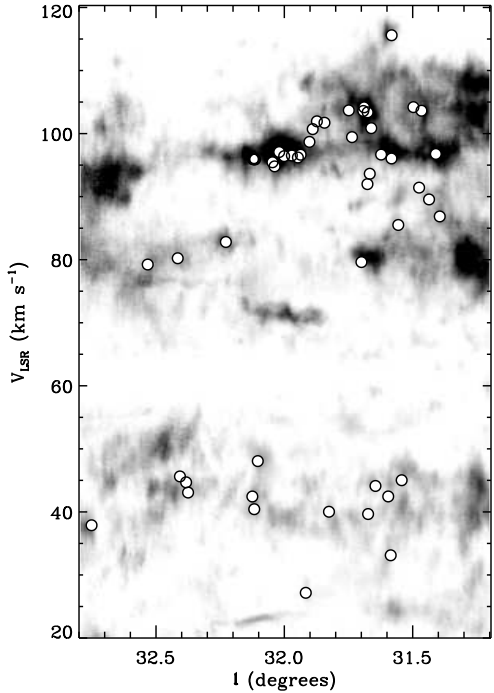}
\caption{\label{l032map} Latitude-integrated $^{13}$CO~($1\to 0$)
  emission with the ($\ell, V$) locations of 42 BGPS/NH$_3$ sources
  indicated.  All BGPS sources are co-located in the
  data cube with significant $^{13}$CO emission, though not
  necessarily the brightest locations.}
\end{figure}

Using the LSR velocities of the BGPS sources, we compare these objects
to the broader distribution of molecular emission in this field.  In
particular, we compare where these sources are found to the
$^{13}$CO~($1\to 0$) data of \citet{bugrs}.  We present a
latitude-integrated map of the molecular emission for the
$\ell=31^{\circ}$ region in Figure \ref{l032map}.  Of note in this
figure, all BGPS sources are associated with the $^{13}$CO emission at
the $>2\sigma$ level.  However, the BGPS sources are not necessarily
found at the brightest positions in the $^{13}$CO data cube.  

Towards these Galactic latitudes, the $^{13}$CO emission is found
predominantly near $V_{\mathrm{LSR}}= 40\mathrm{~km~s}^{-1}$ and
$90\mathrm{~km~s}^{-1}$.  BGPS sources are associated with both
velocity features suggesting a wide range of kinematic distances.  We
explore the association of BGPS emission with the tracers of low
density molecular gas in the full presentation of these results
\citep{bgps-gemob1}.

\begin{figure*}
\plottwo{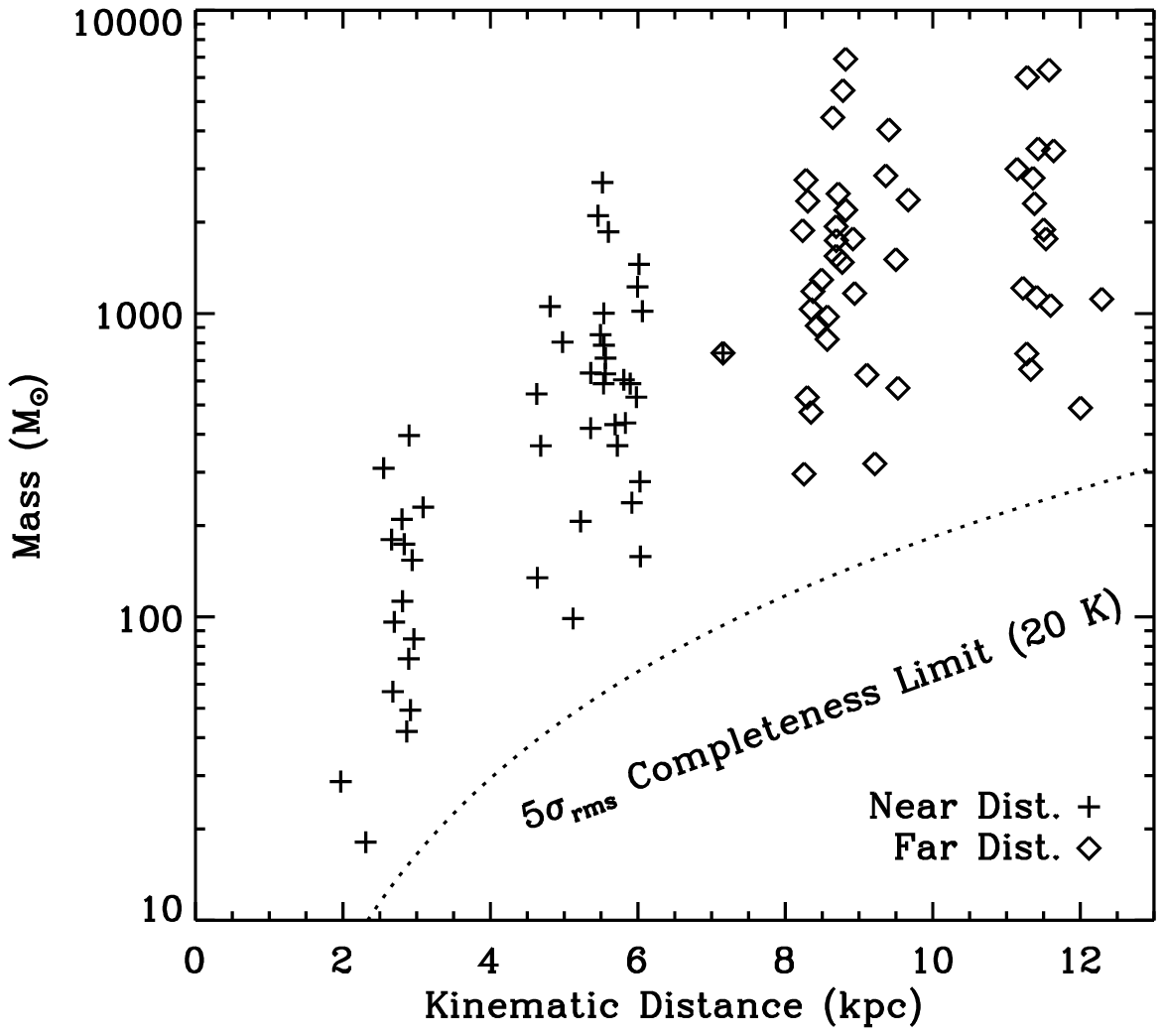}{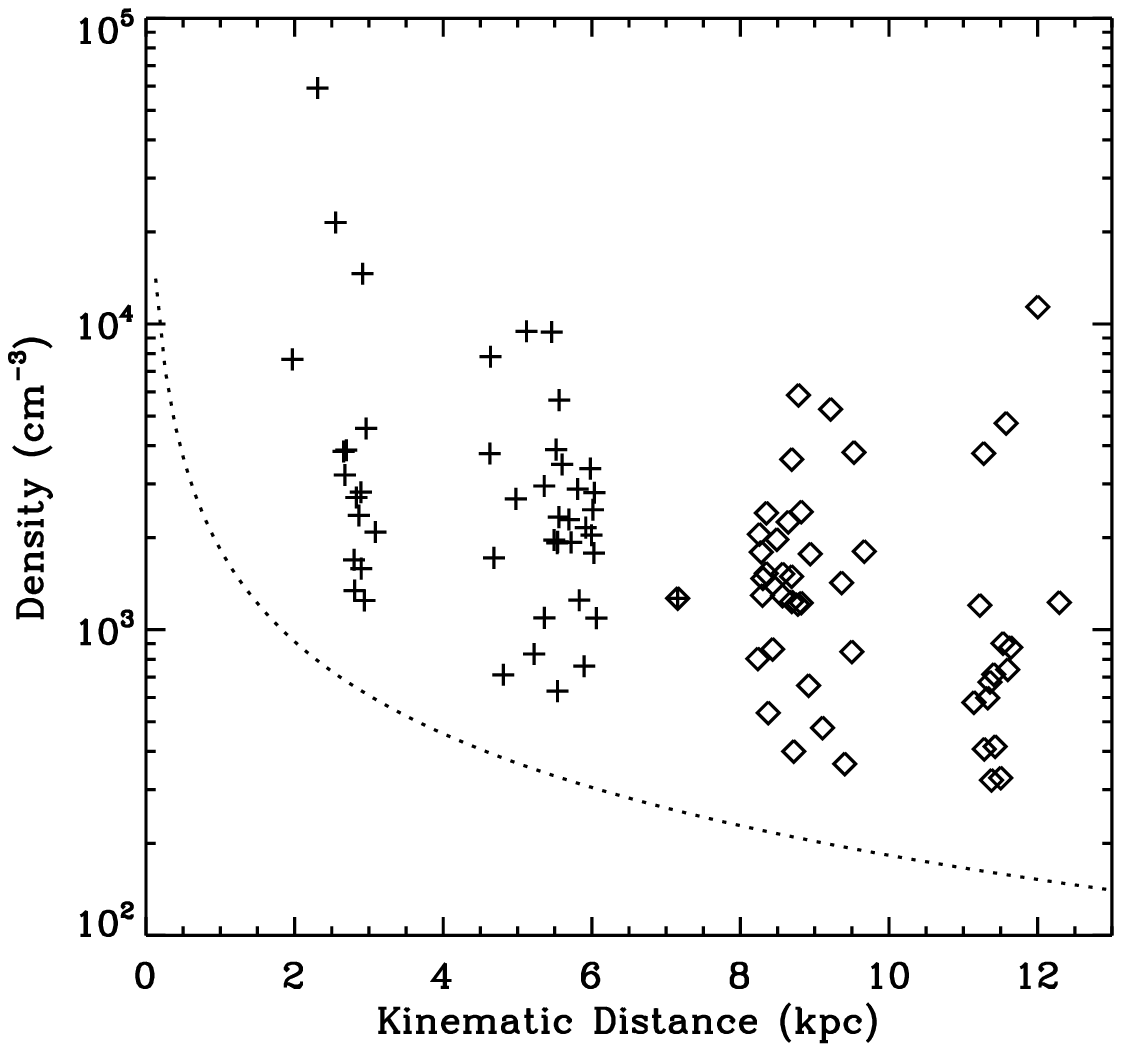}
\caption{\label{sourcemass} Mass estimates for BGPS sources as a
  function of line-of-sight distance (left) and flux density (right).
  The masses are derived from a combination of the BGPS data (for dust
  continuum flux densities) and ammonia data (for gas temperatures and
  kinematic distances) for matched sources in the $\ell=32^{\circ}$
  field.  This figure highlights the wide range in the masses of
  objects recovered in the BGPS catalog.}
\end{figure*}

To explore the properties of BGPS sources in more detail, we estimate
the source masses from the flux density of the continuum emission
combined with distance and temperature estimates from the GBT data:
\begin{eqnarray}
M&=& \frac{d^2 S}{B_\nu (T) \kappa_\nu}\\
&=& 13.1~M_{\odot}~\left(\frac{d}{\mbox{1 kpc}}\right)^2
\left(\frac{S_\nu}{\mbox{1 Jy}} \right)
\left[\frac{\exp(13.0\mbox{ K}/T)-1}{\exp(13.0/20)-1}\right]
\end{eqnarray}
Here, we use the total flux density in the object $S$ measured in the
catalog.  This estimate uses a $T=20$~K dust temperature and
averages the emergent spectrum across the Bolocam band assuming the
dust emissivity scales like $\lambda^{-1.8}$ and an opacity of
$\kappa =0.0114\mathrm{~cm^2~g^{-1}}$ \citep{bolocam-perseus}.  For
these assumptions, the mean frequency of the Bolocam passband is $\nu
=271.1$ GHz.  We plot the masses of the BGPS sources identified as
a function of line-of-sight distance in Figure \ref{sourcemass}.  For
source temperatures, we use the kinetic gas temperatures derived from
the ammonia observations.  We use kinematic distances from the LSR
velocity in the ammonia observations.  Towards these Galactic
longitudes, the velocity-distance relation is double valued resulting
in the distance ambiguity.  We plot the mass and distances derived for
each source at both possible distances; so each point appears twice in
the plot, once on each side of the $d=7.1~$kpc line.  This 7.1 kpc
distance is determined by the tangent velocity at these longitudes.
We also plot a curve representing the derived mass of the $5\sigma$
completeness limit for point sources in this field (where
$5\sigma=0.14 \mbox{ Jy beam}^{-1}$) with a temperature of $T=20$~K.
We note that sources are not distributed uniformly down to the
completeness limit since the ammonia observing targets were selected
by eye and preferentially selected bright objects.

Figure \ref{sourcemass} illustrates the wide range in possible masses
extracted for the sources extracted from the survey data.  First,
depending on the resolution of the kinematic distance ambiguity, the
masses in this field can vary by nearly an order of magnitude.  Even
restricting to near distances, the mass range of detected objects
spans over two orders of magnitude.  While a 15 $M_\odot$ object at
the low mass range may indeed be a high mass {\it core}, the larger
mass scales at farther distances are more appropriately linked to {\it
  clumps}, those large structures which form stellar clusters
\citep{wbm-ppiv}.  This range is largely the result of the wide range
of object flux densities recovered from the original maps (see the
right-hand panel of Figure \ref{sourcemass}), but naturally also
depends strongly on the true distance to the source.

\begin{figure*}
\plottwo{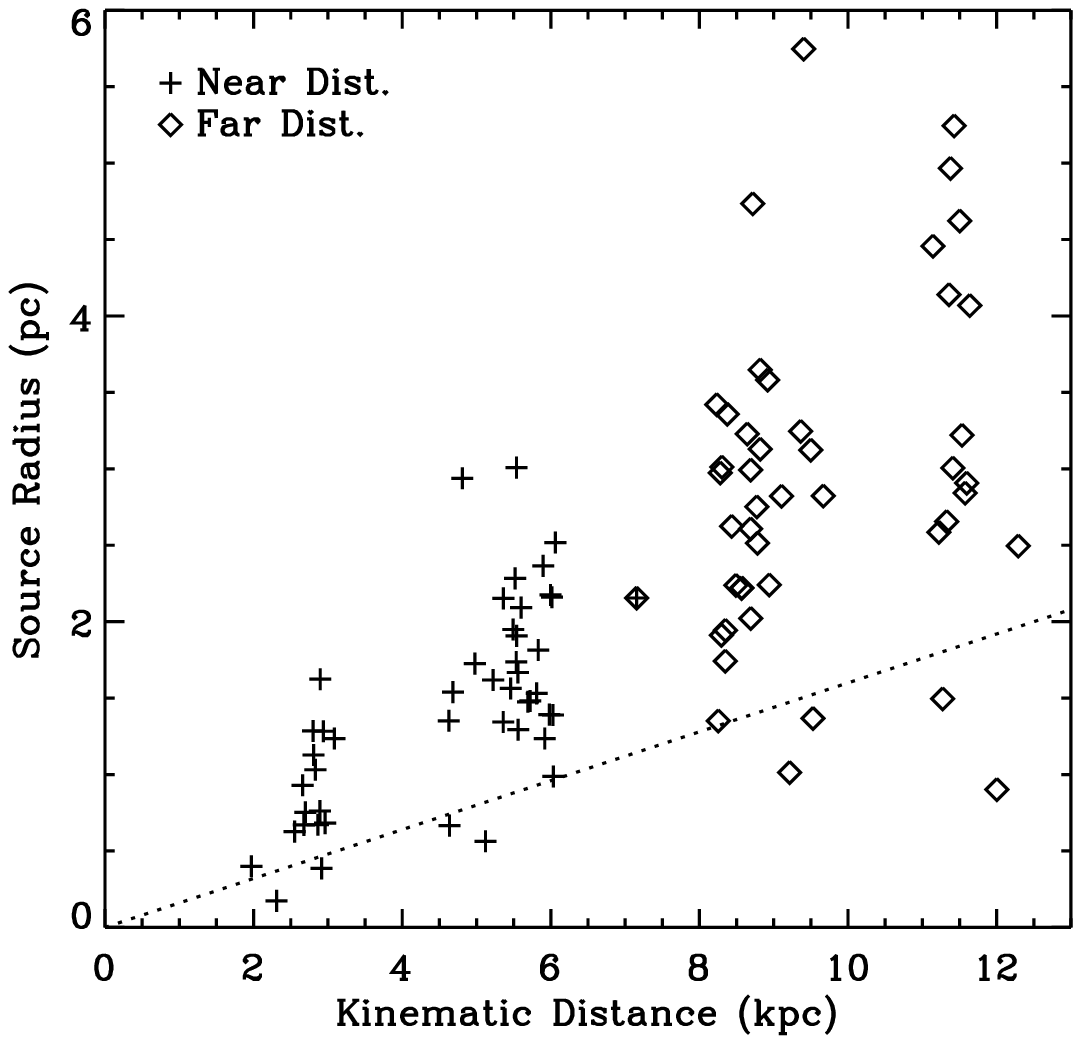}{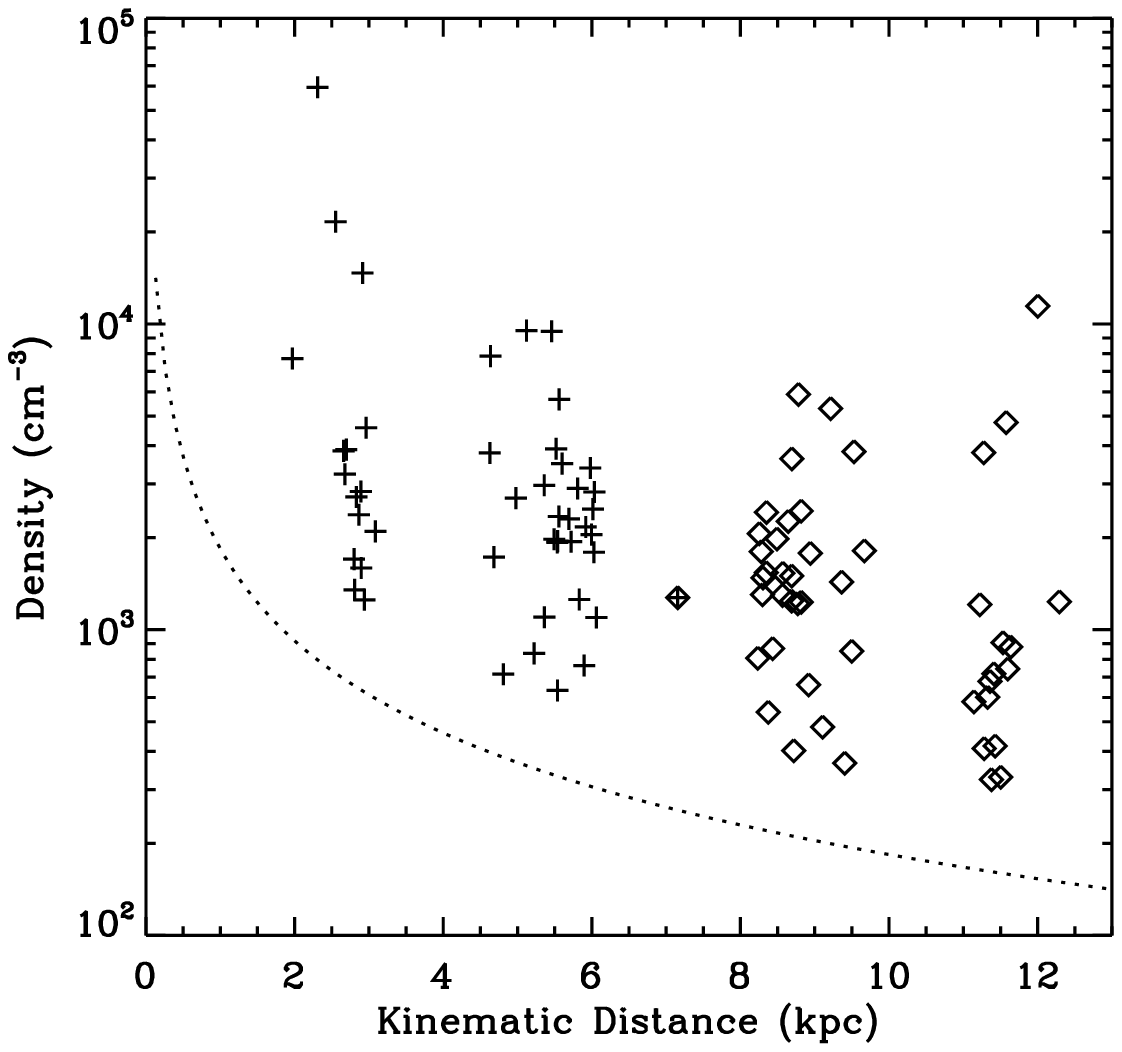}
\caption{\label{raddist} Radius (left) and density (right) estimates
  for BGPS catalog objects matched with ammonia observations in the
  $\ell=32^{\circ}$ field.  Like Figure \ref{sourcemass}, a given
  object is plotted twice for possible near and far kinematic distance
  estimates.  The dotted line indicates the projected size for
  $\theta_R=33''$ (1 beam FWHM) in the left-hand panel.  In the
  right-hand panel, the dotted line indicates the behavior for a
  source with $S=0.15~$Jy =5$\sigma$ and $\theta_R=33''$.  Sources can
  appear below the $\theta_R=33''$ line because the beam deconvolution
  (Equation \ref{radius}).}
\end{figure*}

In Figure \ref{raddist} we plot the derived values of the deconvolved
projected radius $R = d~\theta_R$ and particle density $n_{\mathrm{H}}
= 3M/(4\pi R^3 \mu m_\mathrm{H})$ as functions of distance.  We have
taken the mean particle mass to be $\mu = 2.37$
\citep{kauffmann-mambo}.  In Figures \ref{sourcemass} and
\ref{raddist}, we plot the behavior of sources as a function of
distance to illustrate the importance of distance determinations in
describing the physical properties of these sources.  In these plots,
dotted lines indicate roughly the limiting sizes and densities to
which the survey is sensitive.  Because the ability to estimate the
deconvolved source sizes depends on the signal-to-noise of the
observations, the size limit $\theta_R=33''$(= 1 FWHM) is chosen as a
guide rather than a hard limit.  It is formally possible resolve
sources with sizes smaller than the beam size (Equation \ref{radius})
though the stability of such deconvolutions is suspect for
$\theta_R<\theta_{bm}$.

These plots emphasize the difference in objects recovered by the BGPS
observations of the Galactic plane compared to observations of nearby
molecular clouds where the Bolocam beam and sensitivity is
well-matched to the size of a star-forming core
\citep[e.g.][]{bolocam-oph,bolocam-perseus,enoch07}.  While the BGPS
catalog contains sources with core-like properties ($n\gtrsim
10^{4.5}\mbox{ cm}^{-3}$, $R\lesssim 0.1$~pc), most of the recovered
sources have larger sizes and lower density scales.  The projected
mass and size limits [Figure \ref{sourcemass} (left) and Figure
  \ref{raddist} (left)] limit the detection of individual star-forming
cores in the BGPS to objects within $d\lesssim 1$~kpc.  Beyond this
distance, BGPS is limited to the detection of clumps within larger
clouds.  As is illustrated by Figure \ref{raddist} (right), the
density of gas to which the survey is sensitive, drops off
precipitously with distance to the emitting source.  Objects in the
catalog are thus a range of clumps and cores, and their detection
depends on their distance, physical size, and dust temperature.
Despite these caveats, we note that most sources in this field have
mass and density scales typical of massive, dense clumps in local
clouds.  Given the link that these clumps have to local cluster
formation, it is likely that most sources in the BGPS are associated
with the formation of stellar clusters.

\section{Statistical Properties of BGPS Sources}
\label{summaryprod}
In this section, we describe the basic statistical representations of
the source catalog.  In all cases, the reader is cautioned that the
source catalog is derived from the BGPS maps which are a filtered
representation of the true sky brightness at any given position.  In
particular, large scale emission is filtered out of the final maps on
scales larger than $3.5'$ (Paper I).

\subsection{The Galactic Distribution of Flux Density}

\begin{figure*}
\plotone{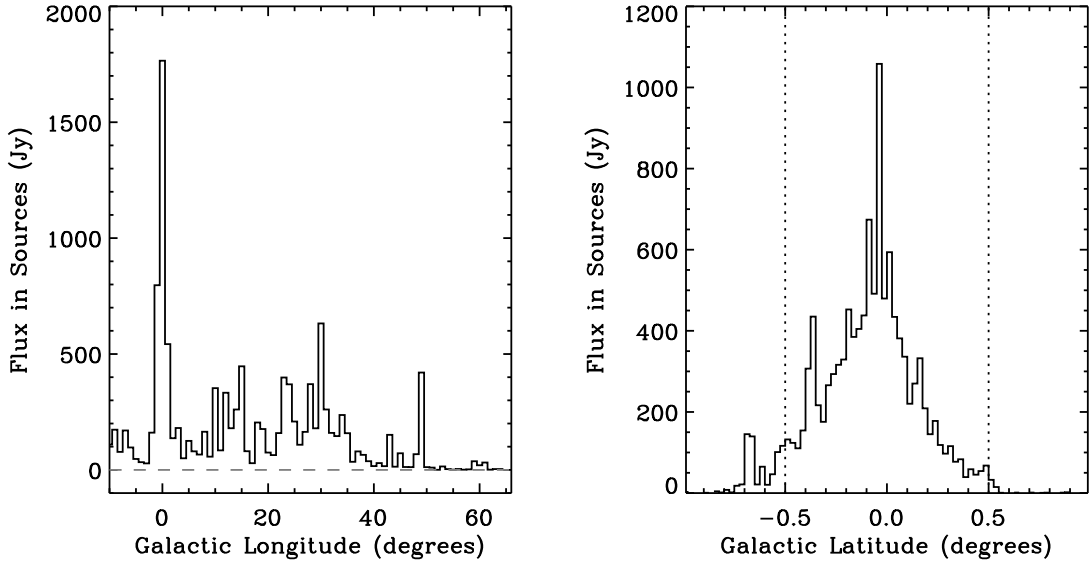}
\caption{\label{lonlat} Longitude and latitude distribution of total
  flux density in catalog sources as a function of longitude and
  latitude in the Galactic plane. The distributions contain sources
  extracted in the $-10^{\circ} < \ell < 66^{\circ}$ region.  For
  $\ell>66^{\circ}$, there is minimal emission found in the map and
  these regions are omitted from the figure.  Dotted lines in the
  right hand panel indicate the rough extent of complete coverage in
  the latitude direction ($\pm 0.5^{\circ}$).}
\end{figure*}

Figure \ref{lonlat} presents the total flux density extracted from the
images in catalog sources as a function of coordinates in the Galactic
plane.  In total, the catalog sources contain 11.9 kJy of flux
density.  Difficulties in recovery of large scale structure
notwithstanding, this represents the emission profile of the Galaxy
along the longitude and latitude direction.

\begin{figure}
\plotone{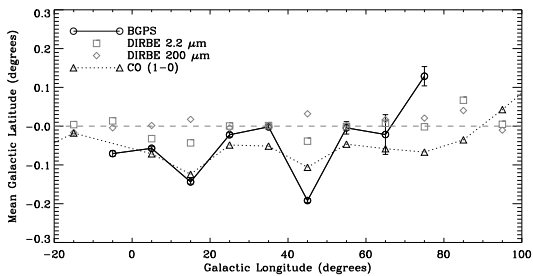}
\caption{\label{meanb} The mean Galactic latitude for various tracers
  of the disk.  The BGPS catalog sources show a mean latitude of
  $\langle b \rangle=-0.095^{\circ}$ which varies significantly with Galactic
  longitude. Data are averaged over the range
  $-0.5^{\circ}<b<0.5^{\circ}$ in 10$^{\circ}$ bins in longitude.  The
  BGPS sources are most closely associated with the CO emission
  whereas tracers of the stellar population (DIRBE 2.2 $\mu$m) and
  warm dust (DIRBE 200 $\mu$m) show less offset towards negative
  latitudes.}
\end{figure}

The right-hand panel of Figure \ref{lonlat} shows that the
distribution of flux density in sources peaks at the flux-weighted
mean $\langle b \rangle =-0.095\pm 0.001$ degrees.  Such an offset was
also seen in the ATLASGAL survey of the $12^{\circ}<b<-30^{\circ}$
region of the Galactic plane at $\lambda$=870 $\mu$m \citep{atlasgal}.
They speculated that the offset may result from the Sun being located
slightly above the Galactic plane.  To further investigate this
possibility, we examined the latitudinal flux density distribution of
BGPS sources as a function of longitude and compared these results to
other tracers of the Galactic disk.  In particular, we compared the
flux density weighted values of the mean latitude for objects in our
catalog to the COBE/DIRBE data at $\lambda=2.2~\mu$m and $200~\mu$m
\citep{dirbe} and the integrated $^{12}$CO (1-0) data of
\citet{dht01}.  The results of the comparison are shown in Figure
\ref{meanb}.  For all data sets, we only compute the flux-density
weighted average latitude over the range $-0.5^{\circ}<b<0.5^{\circ}$
which is the latitude range over which the BGPS is spatially complete.
The data are averaged in 10$^{\circ}$ bins.  The offset of BGPS
sources appears at nearly all Galactic longitudes, but is most
pronounced towards the Galactic center ($-10^{\circ} <\ell<
25^{\circ}$).  The CO (1-0) data follow a similar trend with a
significant offset towards negative latitudes.  We verified that the
effects persist in the CO data if the averaging range is expanded to
larger latitudes (foreground emission from the Aquila rift affects the
averaging for large latitude ranges $|b|<5^{\circ}$).  We also note
that the emission profiles from the DIRBE data do not follow the
offset trends seen in the millimeter continuum or the CO.  The DIRBE
2.2 $\mu$m data primarily traces emission from stellar photospheres
and is affected by dust extinction and the 200 $\mu$m traces warm dust
emission.  Both of these emission features show a more symmetrical
distribution around the Galactic midplane than do those tracers
associated with the molecular gas.  We conclude that the offsets
towards negative latitude are primarily associated with the molecular
ISM.  The significant variations with the mean offset are likely the
result of individual star forming structures rather than a global
property of all Galactic components.  For example, the large negative
excursion in the $40^{\circ}<\ell<50^{\circ}$ bin is due to the W51
star formation complex at ($49.5^{\circ},-0.4^{\circ}$).  At $\ell >
60^{\circ}$, the space density of BGPS sources drops sharply, so the
effects are dominated by the particular distribution of the dense gas
within the molecular complex.  Specifically, in the $\ell \sim
70^{\circ}$ region, there are more dense gas structures at
$b>0^{\circ}$ in the Cygnus X and Cygnus OB 7 region than there are at
$b<0^{\circ}$, while the lower density molecular gas does not exhibit
this small-number effect.

\subsection{The Flux Density Distribution of BGPS Sources}
\label{fluxes}
\begin{figure*}
\plottwo{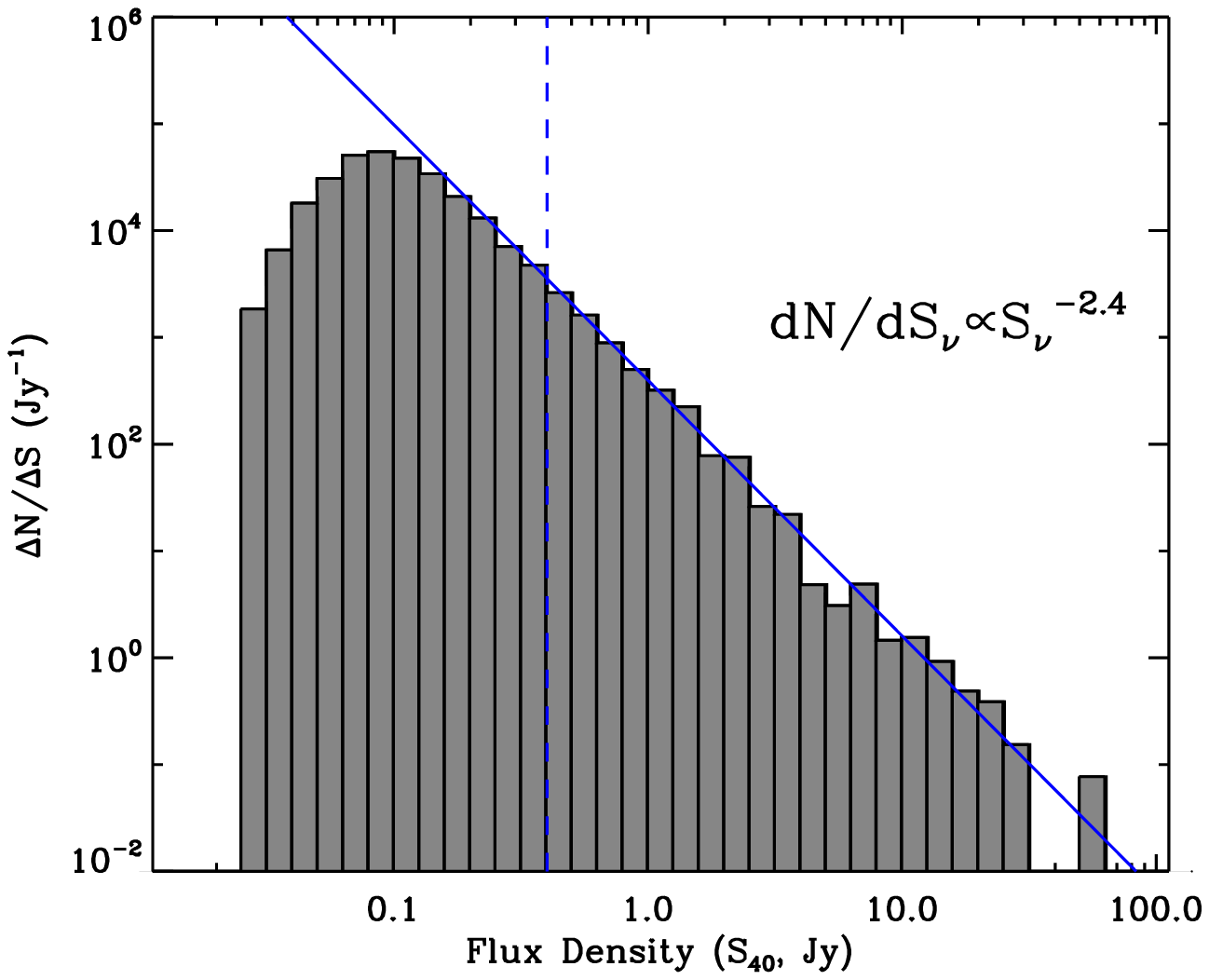}{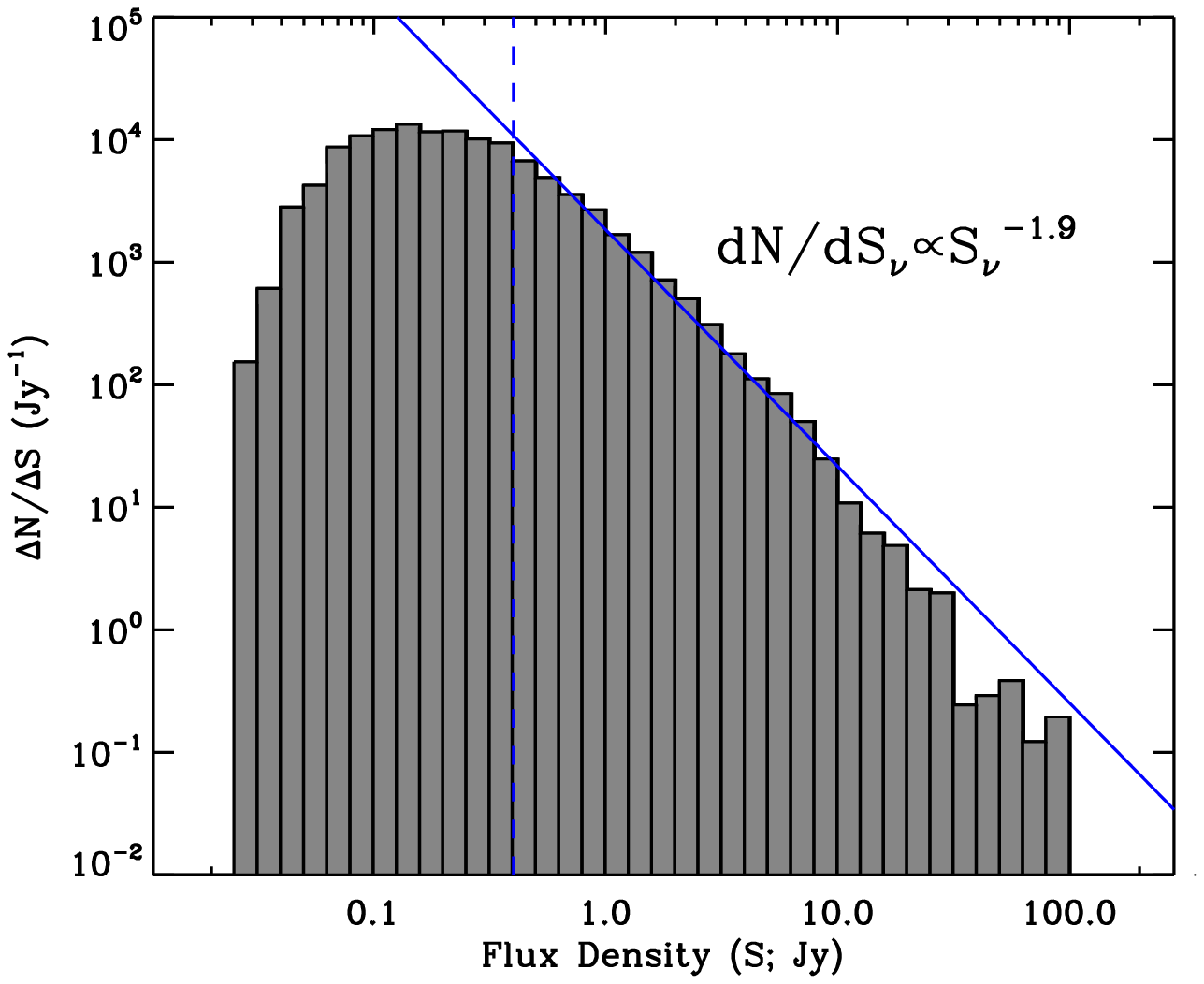}
\caption{\label{fluxdist} Flux density number distribution for objects
  identified in the BGPS.  The left-hand panel shows the flux density
  distribution for $40''$ apertures ($S_{40}$) while the right-hand
  panel shows the total flux density associated with objects ($S$).
  We estimate the power law that best fits each distribution for flux
  densities above the peak in the distribution. Note that the source
  distributions shown are estimators of $dN/dS$, not $dN/d(\log~M)$,
  although the bin widths used to derive the distributions vary
  logarithmically.}
\end{figure*}

In Figure \ref{fluxdist} we present two representations of the flux
density distribution for objects in the BGPS catalog.  The flux
density distribution in $40''$ apertures is shown in the left hand
panel and follows a power-law form over nearly three orders of
magnitude in flux density from the point source completeness limit to
the upper flux density limit in the survey.  We have fit a power law
to the approximation to the differential flux density spectrum $dN/dS
\approx \Delta N/\Delta S \propto S^{-\alpha}$ and find that an index
$\alpha=2.4\pm 0.1$ represents the data quite well.  Despite the
similarity to the Salpeter \citet{salpeter} index $\alpha=2.35$, we
suspect that this is coincidence because the BGPS sources are found
at a wide variety of distances.  This range of distances necessitates
a complete treatment of the conversion of flux density to mass.
Indeed, the Balloon-borne Large Aperture Submillimeter Telescope
(BLAST) survey of the Vela molecular complex at $\lambda = 250,350$
and $500\mu$m found a marginally steeper distribution of object
mass -- $dN/dM\propto M^{-2.8\pm0.2}$ for cold cores -- than the BGPS finds
for flux density \citep{blast-vela}.  A careful treatment of the
distances to BGPS sources may find the mass distribution to be
comparable between BGPS objects and those found in the BLAST survey.

In the right-panel of Figure \ref{fluxdist}, we plot the differential
flux density distribution for the total flux density in the objects
$S$ (Equation \ref{fluxobj}) and find that a power-law is not as good
a representation for the catalog.  Even though the completeness limits
should be the same for both flux density measurements, the power-law
behavior for the total flux density only become apparent above a much
higher flux density level of 0.7 Jy where $\alpha = 1.9\pm 0.1$.  The
difference between these two limits arises entirely because of the
extended structure accounted for in the total flux density
measurement.  A detection near the completeness limit is necessarily
similar to a point source.  To detect extended objects with any degree
of fidelity requires a much higher peak value of the flux density so
that connected, extended structure is apparent.  Thus, source with
flux densities near the completeness limit will not sample the full
range of source sizes that brighter objects will.  Hence, the
detection of extended structure is only apparent in bright sources
leading to a higher limit.  We also note that a power law index fit to
the distribution above the limit of 0.7 Jy gives a significantly
different value compared to the index for flux densities derived for
40$''$ apertures.  This simply reflects the additional effects of the
included sizes of the sources. In general, sources with bright peaks
(measured in the 40$''$ apertures) also tend to be large so their
total recovered flux density is larger than for smaller sources.  This
effectively makes the distribution more top-heavy, reducing the
magnitude of the exponent.

\subsection{Variations in the Flux Density Distribution}
\begin{figure}
\plotone{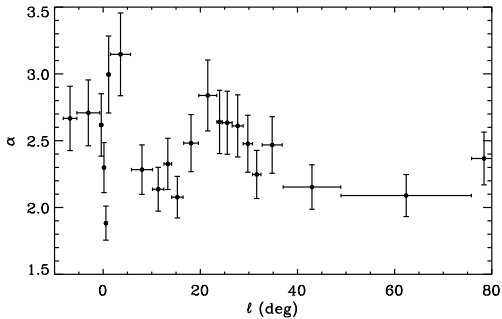}
\caption{\label{alphalon} Power-law index on the flux density
  distribution as a function of Galactic longitude, derived for groups
  of 50 sources above the survey completeness limit of 0.4 Jy.  The
  flux densities are measured in $40''$ apertures across the Galactic
  plane. }
\end{figure}

In Figure \ref{alphalon} we show the value derived for the slope of
the flux density distribution as a function of Galactic latitude over
the contiguous region of the BGPS.  We divide the catalog into groups
of 50 sources based on the galactic longitude above the survey
completeness limit of 0.4 Jy.  The variation of $\alpha$ scatters
around the global average of $\alpha=2.4$ with marginally significant
variation as a function of longitude.  The variations are strongest in
regions of high source density suggesting that blending may contribute
to the derived distribution.  Specifically, the algorithm has a native
angular scale over which it searches for local maxima (a box with edge
length of 2 beam FWHM).  Structures tend to be subdivided on scales
near this search scale.  Large, bright objects in crowded regions of
the plane may be subdivided more often than sources in less crowded
regions.  Indeed, in the $\ell=60^{\circ}$ region, where the source
density distribution is lowest, the flux density distribution is the
most shallow.

\subsection{Size Distribution of Sources}
\begin{figure*}
\plottwo{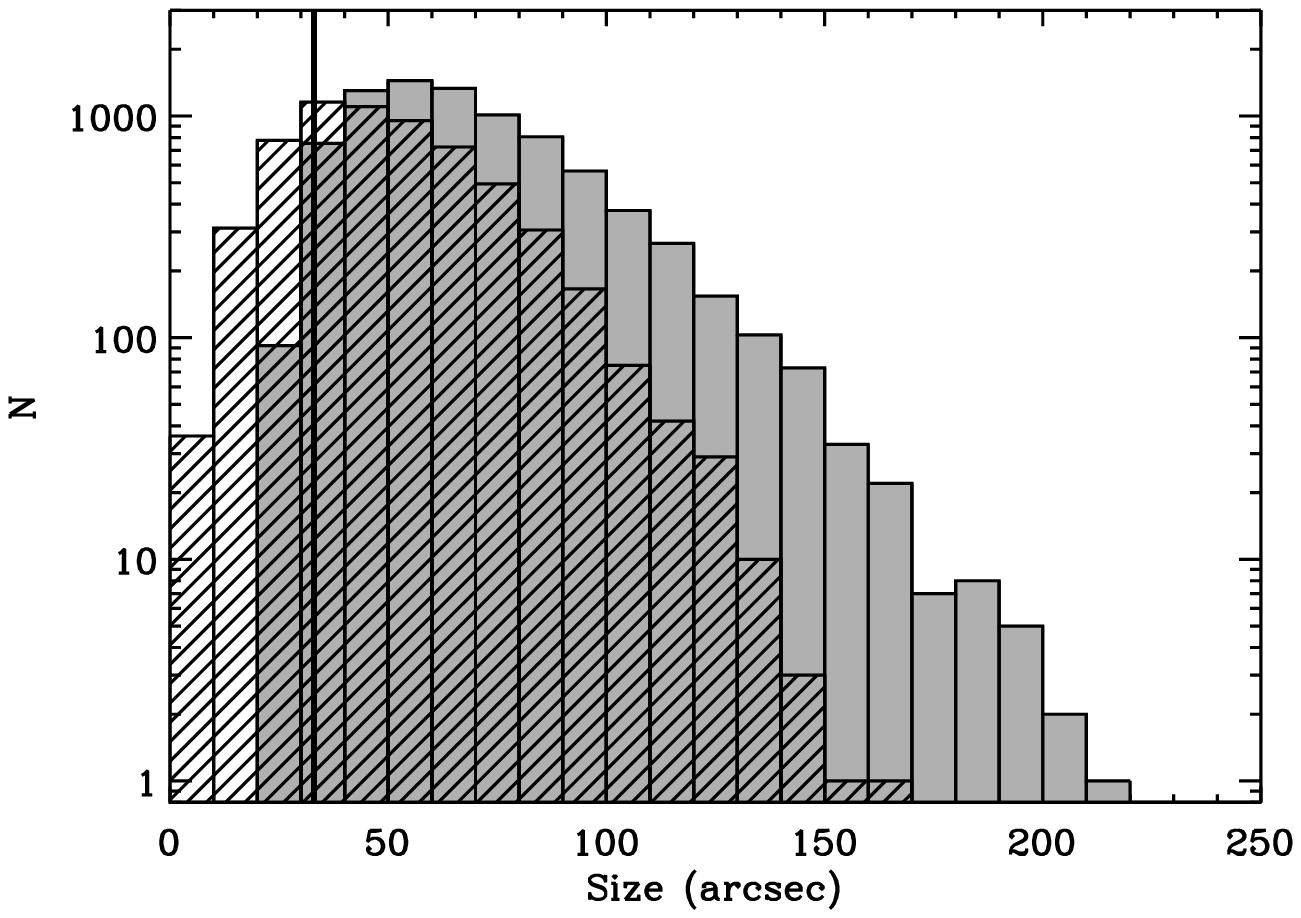}{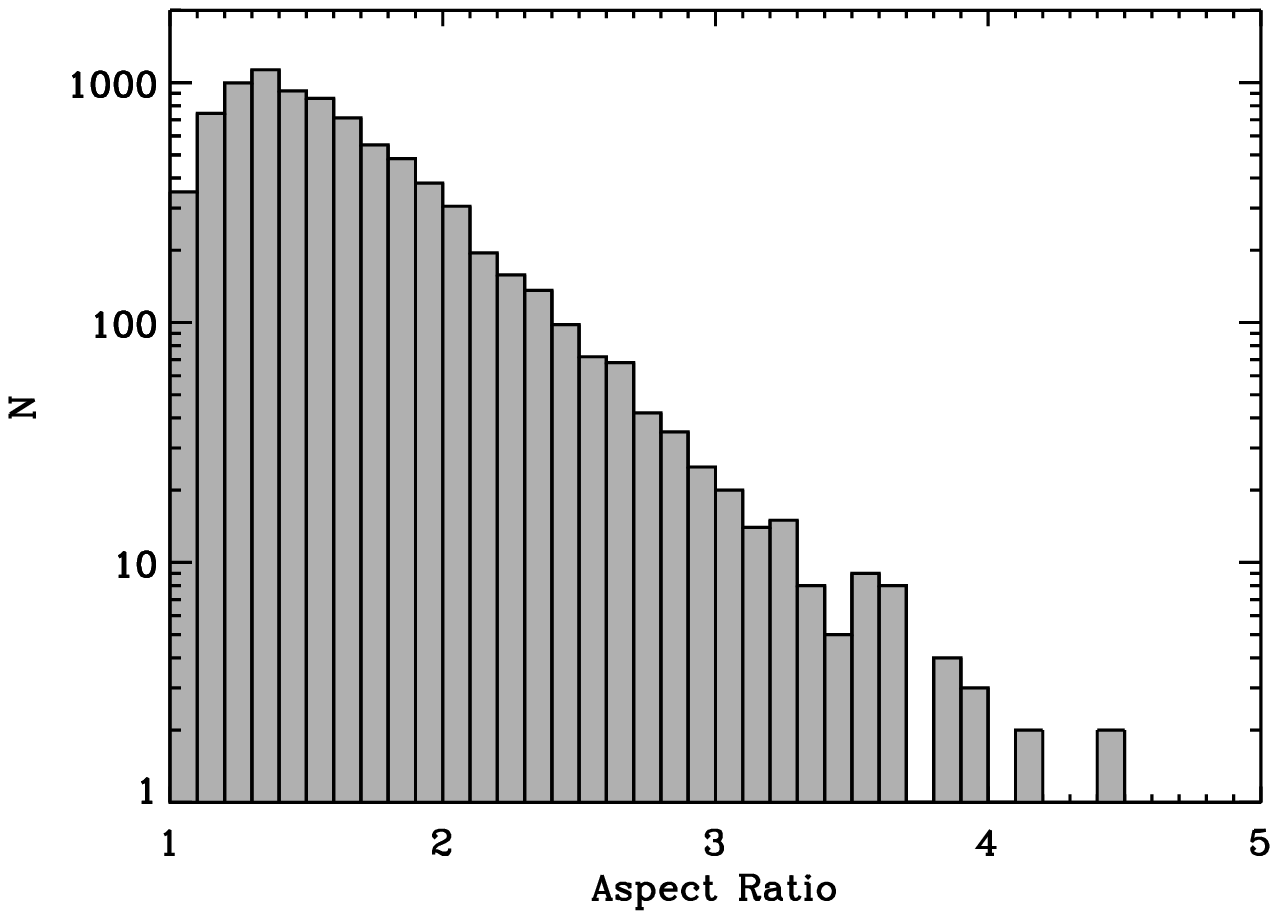}
\caption{\label{dndr} Distributions of source angular sizes ({\it
    left}) and aspect ratios ({\it right}) in the BGPS catalog.  In
  the left panel, the gray shaded histogram indicates the major axis
  size of the sources derived from the moment methods.  The black
  line-filled histogram shows the distribution of the deconvolved
  radii for the sources.  The vertical line is plotted at the FWHM of
  the beam.  In the right panel, the distribution plots the
  major-to-minor aspect ratio for the moments of the source.}
\end{figure*}

The measured sizes of sources varies significantly for the BGPS
sources.  We plot the distribution of source sizes and aspect ratios
in Figure \ref{dndr} for the catalog sources.  We show the major axis
size ($\eta\sigma_{maj}$ where $\eta=2.4$) determined from the moment
method and the deconvolved angular radius of the source ($\theta_R$),
which includes both major and minor axes (Equation \ref{radius}).
Because of the deconvolution, the radius $\theta_R$ can be smaller
than the beam size.  However, we also note that some of the major axis
sizes are also smaller than the beam size.  This results from applying
moment methods without accounting for all emission down to the $I=0$
level.  The clip at $1\sigma$ truncates the low-significance band of
emission around the sources.  For faint sources, the truncation will
clip a significant fraction of the emission resulting in an
underestimate of the source size.  While this effect can be corrected
for, we refrain from doing so since such corrections require
extrapolating the behavior of the sources below the noise floor in the
images.

We also present data for the aspect ratios of the catalog sources
defined as $\sigma_{maj}/\sigma_{min}$.  Both the radii and the aspect
ratios show a peaked distribution with an exponential decline.  The
typical source in the BGPS has a major axis size of $60''$ and a minor
axis size of $45''$.  However, there are significant numbers of
large-aspect-ratio sources which represents asymmetric clump structure
and, for faint sources, the prevalence of filamentary structure in the
survey.

\section{Summary}

We present a catalog of 8358 sources extracted from the Bolocam
Galactic Plane Survey \citep[BGPS, ][]{bgps-paper1}.  The BGPS is a
survey of the Galactic Plane from $-10^{\circ}<\ell < 90.5^{\circ}$
with extensions into the outer Galaxy.  The catalog is generated with
an automated source extraction algorithm, Bolocat, that is designed
for wide-field Bolocam mapping.  The algorithm mimics a by-eye
identification of sources in the BGPS.  The catalog products are
formulated to allow easy follow-up since many of the objects in the
BGPS are relatively unexplored.

The Bolocat algorithm identifies sources based on their significance
with respect to a local estimate of the noise in the BGPS maps.
Regions with high significance are then subdivided into individual
sources based on the presence of local maxima within the region.  Each
pixel in the BGPS image is assigned to at most one catalog source
using a seeded watershed, similar to the Clumpfind \citep{clumpfind}
or SExtractor algorithms \citep{sextractor}.  The properties of the
BGPS sources are measured using the moments of the images for each of
the assigned sources.

We have conducted tests of the Bolocat algorithm using artificial
sources injected into observations in the survey fields.  The Bolocat
algorithm extracts sources with a 99\% completeness limit of
$5\sigma$ where $\sigma$ is the local RMS of the noise.
Since the noise varies across the BGPS fields, the 98\% completeness
limit for the survey is 0.4 Jy, set by the highest noise fields.

Using spectroscopic observations of the NH$_3$(1,1) inversion
transition, we have characterized several of the BGPS sources near
$\ell=32^{\circ}$.  We find that the objects in the catalog are likely
best described as {\it clumps} \citep{wbm-ppiv}, though there is
substantial variation in the properties of the sources.  Since the
catalog is generated using a significance threshold, any source that
produces a signal $>5\sigma$ in the $33''$ beam size of the instrument
will be detected.  Hence, both nearby, small, low-mass objects and
distant, large, high-mass objects can be detected with similar flux
densities.

We find the flux density distribution of sources in the BGPS follows a
power-law form.  The flux densities of objects extracted in a $40''$
aperture follows a power-law over nearly three orders of magnitude:
$dN/dS_{40}\propto S_{40}^{-2.4\pm 0.1}$.  We also find that the mean
Galactic latitude of objects in the survey lies at $\langle b \rangle
=-0.095\pm 0.001^{\circ}$, with some variation across the survey region according
to the presence of large star-forming complexes.

\acknowledgements{The BGPS is supported by the National Science
  Foundation through the NSF grant AST-0708403.  ER acknowledges
  partial support from an NSF AAP Fellowship (AST-0502605) and a
  Discovery Grant from NSERC of Canada. NJE and MKN acknowledge
  support from the NSF grant AST-0607793. We acknowledge the cultural
  role and reverence that the summit of Mauna Kea has within the
  Hawaiian community.  We are fortunate to conduct observations from
  this mountain.  The Green Bank Telescope is operated by the National
  Radio Astronomy Observatory.  The National Radio Astronomy
  Observatory is a facility of the National Science Foundation
  operated under cooperative agreement by Associated Universities,
  Inc. }

{\it Facilities:} \facility{CSO (Bolocam) \facility{GBT (K-band/ACS)}}


\end{document}